\documentclass[12pt]{article}

\pdfoutput=1

\usepackage{epsfig}
\usepackage{amsmath}
\usepackage{amssymb}
\usepackage{graphicx}
\usepackage{subfigure}
\usepackage{array}
\usepackage{cite}
\usepackage{color}
\usepackage{authblk}
\usepackage{epstopdf}
\usepackage[top=1.in,left=1.in,right=1.in,bottom=1.in]{geometry}

\normalsize
\newtheorem {thm}{Theorem}[section]

\def\ba{\begin{array}}
\def\ea{\end{array}}
\def\beq{\begin{equation}}
\def\eeq{\end{equation}}
\def\beqs{\begin{equation*}}
\def\eeqs{\end{equation*}}
\def\bea{\begin{eqnarray}}
\def\eea{\end{eqnarray}}
\def\beas{\begin{eqnarray*}}
\def\eeas{\end{eqnarray*}}
\def\bi{\begin{itemize}}
\def\ei{\end{itemize}}

\def\a{\alpha}
\def\g{\gamma}
\def\d{\delta}
\def\e{\varepsilon}

\def\k{\kappa}
\def\l{\lambda}
\def\m{\mu}

\def\r{\rho}

\def\t{\tau}

\def\o{\omega}
\def\f{\phi}

\def\D{\Delta}
\def\G{\Gamma}

\def\O{\Omega}

\def\hR{\mathbb{R}}

\def\({\textnormal{(}}
\def\){\textnormal{)}}
\def\[{[\neg[}
\def\]{]\neg]}

\def\q{\quad}
\def\qq{\qquad}

\def\neg{\negthinspace}

\def\wt{\widetilde}

\def\cd{\cdot}

\def\b1{{\bf 1}}

\begin{document}
\title{Nonlinear waves in a strongly nonlinear resonant granular chain}

\author[1]{{\normalsize Lifeng Liu}}
\author[2]{{\normalsize Guillaume James}}
\author[3,4]{{\normalsize Panayotis Kevrekidis}}
\author[1]{{\normalsize Anna Vainchtein}}
\affil[1]{{\small\emph{Department of Mathematics\\University of Pittsburgh\\Pittsburgh\\Pennsylvania 15260\\USA}}}
\affil[2]{{\small\emph{INRIA Grenoble - Rh\^one-Alpes, Bipop Team-Project,
Inovall\'ee, 655 Avenue de l'Europe, 38334 Saint Ismier Cedex, France.}}}
\affil[3]{{\small\emph{Department of Mathematics and Statistics\\University of Massachusetts\\Amherst\\Massachusetts 01003\\USA}}}
\affil[4]{{\small\emph{Theoretical Division and Center for Nonlinear Studies\\Los Alamos National Laboratory\\Los Alamos\\New Mexico 87545\\USA}}}

%\date{\today}
%\date{\vspace{-5ex}}
% Toggle commenting to test

\maketitle

\begin{abstract}
We explore a recently proposed locally resonant granular system bearing harmonic
internal resonators in a chain of beads interacting via Hertzian elastic contacts.
In this system, we propose the existence of two types of configurations:
(a) small-amplitude periodic traveling waves and (b) dark-breather solutions, i.e.,
exponentially localized, time periodic states mounted on top of a non-vanishing background.
We also identify conditions under which the system admits long-lived bright breather solutions.
Our results are obtained by means
of an asymptotic reduction to a suitably modified version of
the so-called discrete p-Schr{\"o}dinger (DpS) equation, which is
established as controllably approximating the solutions
of the original system for large but finite times (under suitable
assumptions on the solution amplitude and the resonator mass).
The findings are also corroborated by detailed numerical computations.
A remarkable feature distinguishing our results from other
settings where %modulational instability or
dark breathers
are observed is the complete {\it absence} of precompression
in the system, i.e., the absence of a linear spectral band.
\end{abstract}

\section{Introduction}\label{intro}
Granular materials, tightly packed aggregates of particles that deform elastically when in contact with each other, provide a natural setting for the study of nonlinear waves. Under certain assumptions, dynamics of granular crystals is governed by Hertzian contact interactions of the particles~\cite{Nest01,Sen200821,Coste97}. This leads to the emergence of a wide variety of nonlinear waves.
Among them, arguably, the most prototypical ones are
traveling waves~\cite{Nest01,Sen200821,Coste97}, shock
waves~\cite{herbo,molinari} and exponentially localized (in space),
periodic (in time) states that are referred to as discrete
breathers~\cite{Boechler10,chong14,hasan15}.

Discrete breathers constitute a generic excitation that
emerges in a wide variety of systems and has been thoroughly
reviewed~\cite{Aubry1997201,Flach08}.
Discrete breathers can be divided into two distinct types, which are often referred to as \emph{bright} and \emph{dark} breathers. Bright breathers have tails in relative displacement decaying to zero and are known to exist in dimer (or more generally
heterogeneous) granular
chains with precompression~\cite{Boechler10, Hoog12, Theo10}, monatomic granular chains with
defects~\cite{Theo09} (see also~\cite{jobmelo}) and in Hertzian chains with
harmonic onsite potential~\cite{GJ11,JKC13,BDJ13}. Dark breathers are spatially
modulated standing waves whose amplitude is constant at infinity and
vanishes at the center of the chain (see Fig.~\ref{fig:sw_xnyn_w205}, for example). Their existence, stability and bifurcation structures have been studied in a homogeneous granular chain under precompression \cite{Chong13}.
Recently, experimental investigations utilizing laser Doppler
vibrometry have systematically revealed the existence of such
states in damped, driven granular chains in~\cite{chong14}.
However, to the best of our knowledge, dark breathers have
{\it not} been identified in a monatomic granular chain without precompression.

In this work, we focus on a recent, yet already emerging as particularly
interesting, modification of the standard granular chain, namely the
so-called locally resonant granular chain. The latter belongs to a new type of granular
``metamaterial'' that has additional degrees of freedom and exhibits
a very rich nonlinear dynamic behavior. In particular, in these
systems it is possible to engineer tunable band gaps, as well
as to potentially utilize them for shock absorption and vibration mitigation. Such metamaterials have been recently designed and experimentally tested in the form of chains of spherical beads with internal linear resonators inside the primary beads (mass-in-mass chain)  \cite{Bonanomi14}, granular chains with external ring resonators attached to the beads (mass-with-mass chain) \cite{Gantzounis13} (see also~\cite{PGK12}) and woodpile phononic crystals consisting of vertically stacked slender cylindrical rods in orthogonal contact \cite{Kim14}.
An intriguing feature that has already been reported in such
systems is the presence of weakly nonlinear solitary waves
or nanoptera~\cite{kimyang14} (see also \cite{haitao} for more detailed numerical
results).
Under certain conditions, each of these systems can be described by a granular chain with a secondary mass attached to each bead in the chain by a linear spring. The attached linear oscillator has the natural frequency of the internal resonator in the mass-in-mass chain (Fig.~1 in \cite{Bonanomi14}), the piston normal vibration mode of the ring resonator attached to each bead in the mass-with-mass system (Fig.~9 in \cite{Gantzounis13}) or the primary bending vibration mode of the cylindrical rods in the woodpile setup (Fig.~1 in \cite{Kim14}).

One of the particularly appealing characteristics of a
locally resonant granular chain of this type is the fact that it possesses
a number of special case limits that have previously
been studied. More specifically,
in the limit when the ratio of secondary to primary masses tends to zero, our model reduces to the non-resonant, homogeneous granular chain, while at a very large mass ratio and zero initial conditions for the secondary mass the system approaches a model of Newton's cradle \cite{GJ11}, a granular chain with quadratic onsite potential. In \cite{GJ11} (see also~\cite{JKC13,JS14,BDJ13}),
the so-called discrete p-Schr\"odinger (DpS) modulation equation
governing slowly varying small amplitude of oscillations
was derived and used to prove existence of (and numerically
compute) time-periodic traveling wave
solutions and study other periodic solutions such as standing and traveling
breathers.

In the present setting of locally resonant granular crystals,
we explore predominantly two classes of solutions, namely
(a) small-amplitude periodic traveling waves and (b) (dark)
discrete breathers.
To investigate these solutions at finite mass ratio (i.e., away
from the above studied limits), we follow a similar approach and derive generalized modulation equations of the DpS type. We show that these equations capture small-amplitude periodic traveling waves of the system quite well when the mass ratio is below a critical value.
%Traveling waves with small enough amplitude persist for a long time in the numerical simulations and appear to be stable on finite time scales.
We observe that the system admits only trivial exact bright breathers that involve linear oscillations. However, we use the DpS framework to prove that when the mass ratio is sufficiently large, the system has \emph{long-lived} nontrivial bright breathers for suitable initial conditions. We also use solutions of the DpS equations to form initial conditions for numerical computation of dark breather solutions, whose stability and bifurcation structure are examined for different mass ratios. When the breather frequency is above the linear frequency of the resonator but sufficiently close to it, we identify two families
of dark breathers, as is often the case in nonlinear lattice dynamical
systems~\cite{Flach08}.
The dark breather solutions of site-centered type are long-lived and exhibit marginal oscillatory instability. Meanwhile, the bond-centered solutions exhibit real instability.  This can lead to the emergence of steadily traveling dark breathers in the numerical simulations. In addition, we identify period-doubling bifurcations for the bond-centered solutions. The instability of breather solutions is also affected by the mass ratio. In particular, the real instability of the bond-centered breather solutions at a given frequency gradually becomes stronger as the mass ratio increases.

The paper is organized as follows. Sec.~\ref{sec:model} introduces the model, and the generalized DpS equations are derived in Sec.~\ref{sec:derivation}. In Sec.~\ref{sec:derivationrholarge} we show that for sufficiently large mass ratio the equations reduce to the DpS equation derived in \cite{GJ11} and rigorously justify the validity of this equation on the long-time scale for suitable small-amplitude initial data. We use the modulation equations in Sec.~\ref{sec:tw} to numerically investigate small-amplitude time-periodic traveling waves, including their stability, the accuracy of their DpS approximation and the effect of the mass ratio. In Sec.~\ref{sec:br} we show that the system admits only trivial exact bright breather solutions that do not involve Hertzian interactions. We then prove and numerically demonstrate the existence of long-lived nontrivial bright breathers at sufficiently large mass ratio. In Sec.~\ref{sec:db_approx} we construct the approximate dark breather solutions using the DpS equations. We use these solutions and a continuation procedure based on Newton-type method to compute numerically exact dark breathers and examine their stability and bifurcation structure in Sec.~\ref{sec:db_exact}. Concluding remarks can be found in Sec.~\ref{sec:end}.

\section{The model}\label{sec:model}

Consider a chain of identical particles of mass $m_1$ and suppose a secondary particle of mass $m_2$ is attached to each primary one via a linear spring of stiffness $K>0$ and constrained to move in the horizontal direction. As mentioned in the Introduction, the harmonic oscillator is meant to represent the primary vibration mode of a ring resonator attached to each primary mass or a cylindrical rod. Let $\tilde{u}_n(\tilde{t})$ and $\tilde{v}_n(\tilde{t})$ denote the displacements of the $n$th primary and secondary masses, respectively. The dynamics of the resulting locally resonant granular chain is governed by
\beq
\begin{split}
m_1\dfrac{d^2 \tilde{u}_{n}}{d\tilde{t}^2} &= \mathcal{A}(\tilde{u}_{n-1} - \tilde{u}_n )_{+}^{\a} - \mathcal{A}(\tilde{u}_n - \tilde{u}_{n+1})_{+}^{\a}
- K(\tilde{u}_n - \tilde{v}_n),\\
m_2\dfrac{d^2 \tilde{v}_{n}}{d\tilde{t}^2}&= K(\tilde{u}_{n} - \tilde{v}_{n}).
\label{eq:dyn}
\end{split}
\eeq
 Here $\mathcal{A}(\tilde{u}_n - \tilde{u}_{n+1})_{+}^{\a}$ is the Hertzian contact interaction force between $n$th and $(n+1)$th particles, where $(x)_+ = x$ when $x>0$ and equals zero otherwise, so the particles interact only when they are in contact, $\mathcal{A}>0$ is the Hertzian constant, which depends on the material properties of the contacting particles and radius of the contact curvature, and
 $\a$ is the nonlinear exponent of the contact interaction that depends on the shape of the particles and the mode of contact (e.g. $\a = 3/2$ for spherical beads and orthogonally stacked cylinders).
Typically, we find $\a > 1$, although settings with
$\a < 1$ have also been proposed; see e.g.~\cite{origami} and references
therein.
In writing \eqref{eq:dyn} we assume that the deformation of the particles in contact is confined to a sufficiently small region near the contact point and varies slowly enough on the time scale of interest, so that the static Hertzian law still holds \cite{Coste97};
this is known to be a well justified approximation in a variety of
different settings~\cite{Nest01,Sen200821}. We also assume that dissipation and plastic deformation are negligible, which is generally a reasonable
approximation, although dissipation effects have been argued to potentially
lead to intriguing features in their own right, including
secondary waves~\cite{nester_katja} (see also~\cite{ricardo}).
%valid, for example, for steel beads \cite{Coste97}, and neglect friction.
Choosing $R$ to be a characteristic length scale, for example, the radius of spherical or cylindrical particles, we can introduce dimensionless variables
 $$
 u_n=\dfrac{\tilde{u}_n}{R}, \quad v_n=\dfrac{\tilde{v}_n}{R}, \quad t=\tilde{t}\sqrt{\dfrac{R^{\a-1}\mathcal{A}}{m_1}}
 $$
 and two dimensionless parameters
 $$
 \r=\dfrac{m_2}{m_1}, \quad \k=\dfrac{K}{\mathcal{A}R^{\a-1}},
 $$
 where $\r$ is the ratio of two masses and $\k$ measures the relative strength of the linear elastic spring.
In the dimensionless variables the equations \eqref{eq:dyn} become
\beq
\begin{split}
\ddot u_{n} &= (u_{n-1} - u_n )_{+}^{\a} - (u_n - u_{n+1})_{+}^{\a} - \k(u_n - v_n)\\
\r\ddot v_{n}&= \k(u_{n} - v_{n}),
\label{eq:Hertz}
\end{split}
\eeq
where $\ddot u_{n}$ and $\ddot v_{n}$ are second time derivatives.
In what follows, it will be sometimes convenient to consider \eqref{eq:Hertz} rewritten in terms of relative displacement (strain) variables $x_n = u_n - u_{n-1}$ and $y_n = v_n - v_{n-1}$:
\beq
\begin{split}
\ddot x_{n} &= 2(- x_n )_{+}^{\a} - (- x_{n+1})_{+}^{\a} - (- x_{n-1})_{+}^{\a}  - \k(x_n - y_n)\\
\r\ddot y_{n}&= \k(x_{n} - y_{n}).
\label{eq:HertzS}
\end{split}
\eeq

Note that in the limit $\rho \rightarrow 0$, the model reduces to the one for a regular (non-resonant) homogeneous granular chain. Meanwhile, at $\rho \rightarrow \infty$ and zero initial conditions for $v_n(t)$ the system approaches a model of Newton's cradle, a granular chain with quadratic onsite potential, which is governed by \cite{Hutzler04,GJ11}
\begin{equation}
\label{eq:NC}
\ddot u_n + \k u_n = (u_{n-1} - u_n )_{+}^{\a} - (u_n - u_{n+1})_{+}^{\a}.
\end{equation}
In \cite{GJ11}, the discrete p-Schr\"odinger (DpS) equation
\beq\label{eq:DpS_old}
2i\t_0 \frac{\partial A_n}{\partial\t} = (A_{n+1} - A_{n})|A_{n+1} - A_n|^{\a -1} - (A_n - A_{n-1})|A_{n} - A_{n-1}|^{\a-1}
\eeq
has been derived at $\k=1$ to capture the modulation of small-amplitude nearly harmonic oscillations in the form
\beq\label{eq:approx_zn}
u^{app}_n(t) = \e(A_n(\t)e^{it} + \bar A_n(\t)e^{-it}), \q \t = \e^{\a-1}t,
\eeq
where $\e > 0$ is a small parameter, $A_n(\t)$ is a slowly varying amplitude of the oscillations and $\tau_0$ is a constant depending on $\alpha$. In the next section, we follow a similar approach and use multiscale expansion to derive generalized modulation equations of the DpS type for \eqref{eq:Hertz} with finite $\rho$.

%%%%%%%%%%%%%%%%% sec: Modulation equations for small amplitude waves %%%%%%%%%%%%%%%%
\section{Modulation equations for small amplitude waves}\label{sec:modulation}
\subsection{Derivation of generalized DpS equations at finite $\rho$}\label{sec:derivation}
Using the two-timing asymptotic expansion as in \cite{GJ11}, we seek solutions of \eqref{eq:Hertz} in the form
\beqs
u(t)=\e U(t,\t), \q v(t)=\e V(t,\t)
\eeqs
where $\t=\e^{\a-1}t$ is the slow time, and $u$, $v$, $U$ and $V$ are vectors with components $u_n$, $v_n$, $U_n$, $V_n$, respectively.
The governing equations \eqref{eq:Hertz} then yield
\beq
\begin{split}
&(\partial_t+\e^{\alpha-1}\partial_\tau)^2 U=\e^{\alpha-1} G(U)+\k(V-U)\\
&(\partial_t+\e^{\alpha-1}\partial_\tau)^2 V=\dfrac{\k}{\rho}(U-V),
\end{split}
\label{eq:full}
\eeq
where the nonlinear term is given by
\beqs
G(U)_n=(U_{n-1}-U_n)^\alpha_{+}-(U_n-U_{n+1})^\alpha_{+}.
\eeqs
The solution has the form
\beqs
U=U^0+\e^{\a-1}U^1+o(\e^{\a-1}), \quad V=V^0+\e^{\alpha-1}V^1+o(\e^{\a-1}).
\eeqs
The $0$th order terms satisfy a linear system, which after the elimination of secular terms yields
\beq
U^0=B(\tau)+\k[A(\tau)e^{i\omega t}+\bar{A}(\tau)e^{-i\omega t}], \quad V^0=B(\tau)-\dfrac{\k}{\rho}[A(\tau)e^{i\omega t}+\bar{A}(\tau)e^{-i\omega t}],
\label{eq:U0V0}
\eeq
where $\o = \sqrt{\k + \k/\r}$ is the frequency of harmonic oscillations.
This internal frequency of each resonator
is associated with the out-of-phase motion of the
displacements $U$ and $V$.
On the order $O(\e^{\alpha-1})$, the system \eqref{eq:full} results in
\beq
\begin{split}
&(\partial_t^2+\k)U^1-\k V^1=-2\k\partial_\tau Ai\omega e^{i\omega t}+c.c.+ G(U^0)\\
&\partial_t^2 V^1-\dfrac{\k}{\rho}(U^1-V^1)=\dfrac{2\k}{\rho}\partial_\tau Ai\omega e^{i\omega t}+c.c.,
\end{split}
\label{eq:higher_order}
\eeq
where $c.c.$ denotes the complex conjugate. Let
\beq
J(f)=\frac{\omega}{2\pi}\int_0^{2\pi/\omega}f(t)e^{-i\omega t}dt
\label{eq:J_def}
\eeq
denote the projection of $f(t)$ on $e^{i\o t}$ and define the averaging operator as
\beq
E(f)=\dfrac{\omega}{2\pi}\int_0^{2\pi/\omega}f(t)dt.
\label{eq:aver_def}
\eeq
The projection operator on all remaining Fourier modes (of the form $e^{ij\omega t}$, $j \neq \pm 1$, $j \neq 0$) is given by
\beq
\Pi_h=I-E-e^{i\omega t}J-e^{-i\omega t}\bar{J}.
\label{eq:Pi_h}
\eeq
Let $U_h^1=\Pi_h U^1$, $V_h^1=\Pi_h V^1$. Then \eqref{eq:higher_order} yields
\beqs
\begin{split}
&(\partial_t^2+\k)U^1_h-\k V^1_h=\Pi_h G(U^0)\\
&\partial_t^2 V^1_h-\dfrac{\k}{\rho}(U^1_h-V^1_h)=0.
\end{split}
\eeqs
Note that this equation has a unique $2\pi/\omega$-periodic solution $(U_h^1,V_h^1)^T$ because for each
$j$ such that $j \neq \pm 1$, $j \neq 0$, the matrix
\beqs
\left[\begin{array}{cc}\k-j^2\omega^2 & -\k\\-\k/\rho & \k/\rho-j^2\omega^2\end{array}\right]
\eeqs
for the left hand side of the above equation associated with $j$th harmonic is invertible. Let
\beqs
U^1=U_h^1+C^0(\tau)+C^1(\tau)e^{i\omega t}+c.c., \quad
V^1=V_h^1+D^0(\tau)+D^1(\tau)e^{i\omega t}+c.c.
\eeqs
and project \eqref{eq:higher_order} on $e^{i\omega t}$, recalling that $\o = \sqrt{\k + \k/\r}$:
\beqs
\begin{split}
&-(\k/\rho)C^1-\k D^1=-2\k\partial_\tau Ai\omega+ J(G(U^0))\\
&-(\k/\rho)C^1-\k D^1=(2\k/\rho)\partial_\tau Ai\omega.
\end{split}
\eeqs
This yields the compatibility condition
\beq
2i\k\omega^3\partial_\tau A=J(G(U^0))
\label{eq:S1}
\eeq
and
\beqs
D^1=-\dfrac{\k}{\rho}(C^1+2i\omega\partial_\tau A).
\eeqs
Taking the average of \eqref{eq:higher_order}, we obtain
\beqs
\begin{split}
&\k(C^0-D^0)=E(G(U^0))\\
&-\dfrac{\k}{\rho}(C^0-D^0)=0.
\end{split}
\eeqs
Since $\rho$ is finite, this yields $C^0=D^0$, and hence the following
condition can be obtained for the leading order solution:
\beq
E(G(U^0))=0.
\label{eq:S2}
\eeq

To obtain the generalized DpS equations, we now consider the conditions \eqref{eq:S1} and \eqref{eq:S2} in more detail. Observe that for $b \in \mathbb{R}$, $z= re^{i\theta} \in \mathbb{C}$, we have
\beq
E[(-b+\k ze^{i\omega t}+\k \bar{z}e^{-i\omega t})_+^\alpha]=\dfrac{1}{2\pi}\int_0^{2\pi}(-b+2\k r\cos t)^\alpha_{+} dt
\equiv g_\alpha(b,r), \quad r=|z|.
\label{eq:aux1}
\eeq
Here we rescaled time in the averaging integral and used the fact that the result is independent of $\theta$ since we can always shift time when averaging. Similarly,
\beq
J[(-b+\k ze^{i\omega t}+\k \bar{z}e^{-i\omega t})_+^\alpha]=\dfrac{z}{2\pi r}\int_0^{2\pi}e^{-it}(-b+2\k r\cos t)^\alpha_{+} dt
\equiv z h_\alpha(b,r), \quad r=|z|.
\label{eq:aux2}
\eeq
Defining the forward and backward shift operators
$$
(\delta^+A)_n=A_{n+1}-A_n, \quad (\delta^-A)_n=A_{n}-A_{n-1},
$$
we observe that
$$
G(U^0)=-\delta^+(-\delta^-U^0)^\alpha_+=-\delta^+(-\delta^-B(\tau)-\k\delta^- A(\tau)e^{i\omega t}-\k\delta^-\bar{A}(\tau)e^{-i\omega t})^\alpha_+,
$$
where we used the first of \eqref{eq:U0V0} to obtain the second equality.
Substituting this in \eqref{eq:S1} and \eqref{eq:S2} and using \eqref{eq:aux1}, \eqref{eq:aux2} with $z=-\delta^-A$ and $b=\delta^-B$, we obtain the \emph{generalized DpS equations}
\beq
2i\k\omega^3\partial_\tau A=\delta^+[h_\alpha(\delta^-B,|\delta^-A|)\delta^-A]
\label{eq:S1new}
\eeq
and
\beq
\delta^+g_\alpha(\delta^-B,|\delta^-A|)=0.
\label{eq:S2new}
\eeq

\subsection{DpS equation at large $\rho$}\label{sec:derivationrholarge}
We now investigate the case of large $\rho$.
Consider first the ``critical" case when $\rho=\e^{1 - \alpha}$. The $0$th order problem is
$$
\partial^2_t U^0=\k(V^0-U^0), \quad \partial^2_tV^0=0,
$$
which yields
\beq
U^0=B(\tau)+\k [A(\tau)e^{i\sqrt{\k} t}+\bar{A}(\tau)e^{-i\sqrt{\k} t}], \quad V^0=B(\tau),
\label{eq:U0V0_large_rho}
\eeq
where we used the fact that $\omega=\sqrt{\k}$ when $\r \rightarrow \infty$. Meanwhile, the $O(\e^{\alpha-1})$ problem becomes
\beqs
\begin{split}
&(\partial_t^2+\k)U^1-\k V^1=-2i\k^{3/2}\partial_\tau A e^{i\sqrt{\k}t}+c.c.+ G(U^0)\\
&\partial_t^2 V^1=\k^2Ae^{i\sqrt{\k}t}+c.c.
\end{split}
\eeqs
Note that the right hand side of the second equation has zero time average, as it should to be consistent with the left hand side, and the equation yields
$$
V^1=-\k A(\tau)e^{i\sqrt{\k}t}+c.c.+D^0(\tau).
$$
Putting this back into the first equation and projecting on $e^{i\sqrt{\k}t}$, we get
$$
2\k^{3/2}i\partial_\tau A+\k^2 A=J[G(U^0)],
$$
which is \emph{almost} like the DpS equation in \cite{GJ11} if we set $\k=1$. Note, however, the additional term $\k^2A$ in the left hand side and the fact that $U^0$ also includes $B(\tau)$. Observe further that there are no
conditions to determine $B$ at this order. If $B=0$, we
get a (modified) DpS equation for $A$ only.

Now suppose $\rho=\e^{1-\g}$, $\g>\alpha$. Then the $0$th order equation is the same, so the solution is still given by \eqref{eq:U0V0_large_rho}, while on $O(\e^{\alpha-1})$ we get
\beqs
\begin{split}
&(\partial_t^2+\k)U^1-\k V^1=-2\k^{3/2}\partial_\tau Ai e^{i\sqrt{\k} t}+c.c.+ G(U^0)\\
&\partial_t^2 V^1=0,
\end{split}
\eeqs
so the second equation yields $V^1=D_0(\tau)$, while the projection of the first on $e^{i\sqrt{\k}t}$ yields the DpS equation for the Newton's cradle model:
$$
2\k^{3/2}i\partial_\tau A=J[G(U^0)].
$$
Note that $B(\tau)$ in \eqref{eq:U0V0_large_rho} is again not determined at this order. Observe, however, that in the limit $\r \rightarrow \infty$ the initial conditions $v(0)=\dot{v}(0)=0$ yield $v(t) \equiv 0$ (and thus $V^0=V^1=0$), and we recover
\eqref{eq:approx_zn} and the DpS equation \eqref{eq:DpS_old} at $\k=1$~:
\begin{equation}
\label{eq:DpSlr}
i\, \partial_\tau A=\omega_0\, \delta^+[|\delta^-A|^{\alpha -1}\delta^-A]
\end{equation}
with (see \cite{GJ11})
\begin{equation}
\label{eq:defom0}
\omega_0 = \frac{2^{\alpha -2}}{\sqrt{\pi}}\frac{\G(\a/2+1)}{\G((\a+1)/2+1)}.
\end{equation}
In Theorem \ref{th:dps} below,
we justify the DpS equation \eqref{eq:DpSlr} on long time scales, for suitable small-amplitude initial conditions.
We obtain error estimates between solutions
of (\ref{eq:Hertz}) and modulated profiles described by DpS.
We seek solutions of (\ref{eq:Hertz}) and the DpS equation
in the usual sequence spaces $\ell_p$ with $1\leq p \leq +\infty$.
For simplicity we state Theorem \ref{th:dps} in the case $\kappa =1$.

\begin{thm}
\label{th:dps}
Fix constants $C_{\rm{r}},C_{\rm{i}} ,T >0$ and a solution $A\in \mathcal{C}^2 ([0,T],\ell_p)$ of the DpS equation \eqref{eq:DpSlr}.
There exist constants $\e_T>0$ and $C_T \geq C_{\rm{i}}$ such that the following holds:\\
For all $\e\leq\e_T$ and for $\rho^{-1} \leq C_{\rm{r}}\, \e^{2(\alpha -1)}$,
for all initial condition $(u(0),v(0),\dot{u}(0),\dot{v}(0)) \in  \ell_p^4$ satisfying
\begin{equation}
\label{condciu}
\| u(0)-2\e \, {\rm{Re}}\, A(0) \|_p + \| \dot{u}(0)+2\e \, {\rm{Im}}\, A(0) \|_p \leq C_{\rm{i}}\e^{\alpha},
\end{equation}
\begin{equation}
\label{condciv}
\| v(0) \|_p \leq C_{\rm{i}}\e^{\alpha}, \ \ \
\| \dot{v}(0) \|_p \leq C_{\rm{i}}\e^{2\alpha -1},
\end{equation}
the corresponding solution of (\ref{eq:Hertz}) satisfies for all $t\in[0,T/\e^{\alpha-1}]$
\begin{equation}
\label{estimu}
\| u(t)-2\e \, {\rm{Re}}\, (A(\e^{\alpha-1}t)\, e^{it}) \|_p
+ \| \dot{u}(t)+2\e \, {\rm{Im}}\, (A(\e^{\alpha-1}t)\, e^{it}) \|_p \leq C_{T}\e^{\alpha},
\end{equation}
\begin{equation}
\label{estimv}
\| v(t) \|_p \leq C_{T}\e^{\alpha}, \ \ \
\| \dot{v}(t) \|_p \leq C_{T}\e^{2\alpha -1}.
\end{equation}
\end{thm}

Similarly to what was established in \cite{BDJ13} for Newton's cradle problem, Theorem \ref{th:dps}
shows that small $O(\e )$ solutions of (\ref{eq:Hertz}) are described by the DpS equation over
long (but finite) times of order $\e^{1-\alpha}$. However, there are important differences compared
to the results of \cite{BDJ13}. Firstly, the DpS approximation is not valid for all small-amplitude initial conditions,
since one has to assume that $v(0)$ and $\dot{v}(0)$ are small enough (see (\ref{condciv})). Secondly, $\rho$ must
be large when $\e$ is small. More precisely,
$\rho$ must be greater than $\e^{2(1-\alpha )}$, which scales as the square of the
characteristic time scale of DpS. This is due to the translational invariance of (\ref{eq:Hertz}), which introduces a
Jordan block in the linearization of (\ref{eq:Hertz}) around the trivial state, inducing a quadratic growth
of secular terms (see the estimate (\ref{estimr1}) below).\\

Let us now prove Theorem \ref{th:dps}. The main steps are
Gronwall estimates to obtain solutions of (\ref{eq:Hertz}) close to solutions of the Newton's cradle problem (\ref{eq:NC}) when $\rho$ is large,
and the use of the results of \cite{BDJ13} to approximate solutions of (\ref{eq:NC}) with the DpS equation.

Equation (\ref{eq:Hertz}) at $\kappa =1$ reads
\begin{eqnarray}
\label{eq:u}
\ddot{u}+u-v &=& G(u), \\
\label{eq:v}
\ddot{v}&=& \sigma \, (u-v),
\end{eqnarray}
where $\sigma = {1}/{\rho}$ is a small parameter satisfying
\begin{equation}
\label{sizemu}
\sigma \leq C_{\rm{r}}\, \e^{2(\alpha -1)},
\end{equation}
as assumed in Theorem \ref{th:dps}.
In addition, we have
\begin{equation}
\label{sizeG}
\| G(u) \|_p = O(\| u\|_p^\alpha), \ \ \
\| DG(u) \|_{\mathcal{L}(\ell_p)} = O(\| u\|_p^{\alpha-1}) .
\end{equation}
To simplify subsequent estimates, it is convenient to uncouple
the linear parts of (\ref{eq:u}) and (\ref{eq:v}),
which can be achieved by making the change of variables
$$
u=Q+R, \ \ \ v = R - \sigma\, Q.
$$
The system (\ref{eq:u})-(\ref{eq:v}) is equivalent to
\begin{eqnarray}
\label{eq:q}
\ddot{Q}+Q &=& \chi\, G(Q+R) - \sigma\, Q, \\
\label{eq:r}
\ddot{R}&=& \chi \sigma \, G(Q+R),
\end{eqnarray}
where $\chi = (1+\sigma )^{-1}$ is close to unity.

To approximate the dynamics of (\ref{eq:q})-(\ref{eq:r}) when $\sigma$ is small, we first
consider the case $\sigma =0 $ and $R=0$ of (\ref{eq:q}) (leading to the Newton's cradle problem)
and use the results of \cite{BDJ13} relating Newton's cradle problem to the DpS equation. More precisely,
given the solution $A$ of the DpS equation considered in Theorem \ref{th:dps}, we introduce
the solution $Q_{\rm{a}}$ of
$$
\ddot{Q}_{\rm{a}}+Q_{\rm{a}} =  G(Q_{\rm{a}})
$$
with initial condition $Q_{\rm{a}}(0)= 2\e \,{\rm{Re}}\, A(0)$, $\dot{Q}_{\rm{a}}(0)= -2\e \,{\rm{Im}}\, A(0)$.
According to Theorem 2.10 of \cite{BDJ13}, for $\e $ small enough, the solution $Q_{\rm{a}}$ is defined on a maximal interval
of existence $(t^- , t^+)$ containing $[0 , T\, \e^{1-\alpha}]$ and satisfies
for all $t\in [0 , T\, \e^{1-\alpha}]$
\begin{equation}
\label{estimerrqa}
\| Q_{\rm{a}}(t)-2\e \, {\rm{Re}}\, (A(\e^{\alpha-1}t)\, e^{it}) \|_p
+ \| \dot{Q}_{\rm{a}}(t)+2\e \, {\rm{Im}}\, (A(\e^{\alpha-1}t)\, e^{it}) \|_p \leq C\, \e^{\alpha}.
\end{equation}
This implies in particular that
\begin{equation}
\label{estimqa}
\| Q_{\rm{a}} \|_{L^\infty ((0,T\, \e^{1-\alpha}),\ell_p)}
+ \| \dot{Q}_{\rm{a}} \|_{L^\infty ((0,T\, \e^{1-\alpha}),\ell_p)} \leq M\, \e.
\end{equation}
Next, our aim is to show that $Q$ remains close to $Q_{\rm{a}}$
(and its DpS approximation) and $R$ remains small over long times,
for suitable initial conditions, $\e$ small enough and $\rho$ large enough
(i.e. $\sigma$ small enough).
Setting $Q=Q_{\rm{a}} + W$ in (\ref{eq:q})-(\ref{eq:r}) yields
\begin{eqnarray}
\label{eq:W}
\ddot{W}+W &=& N (W,R) \\
\label{eq:rbis}
\ddot{R}&=& \chi \sigma \, G(Q_{\rm{a}} + W+R),
\end{eqnarray}
where
\begin{equation}
\label{defn}
N (W,R)=-\sigma (  Q_{\rm{a}} + \chi\, G(Q_{\rm{a}})  ) + \chi\, [\, G(Q_{\rm{a}} + W+R)- G(Q_{\rm{a}})  \, ] -\sigma\, W .
\end{equation}
Moreover, from the identities
$$
W=Q-Q_{\rm{a}}=(1+\sigma )^{-1}\, (u-v) -Q_{\rm{a}}, \ \ \
R=(1+\sigma )^{-1}\, (\sigma u + v)
$$
and the assumptions (\ref{condciu})-(\ref{condciv}) and (\ref{sizemu}), it follows that
\begin{equation}
\label{condciz}
\| W(0) \|_p + \| \dot{W}(0) \|_p
\leq C\, \e^{\alpha},
\end{equation}
\begin{equation}
\label{condcivp}
\| R(0) \|_p \leq C\, \e^{\alpha}, \ \ \
\| \dot{R}(0) \|_p \leq C\, \e^{2\alpha -1}.
\end{equation}
Let $X=(W,R,\dot{W},\dot{R})$ and
$\| X \|_p$ denote the sum of the $\ell_p$ norms of each component.
We shall now use Gronwall estimates to bound
$\| X(t) \|_p$ on the time scale considered in Theorem \ref{th:dps}.

The solutions of (\ref{eq:W})-(\ref{eq:rbis}) corresponding to initial conditions satisfying (\ref{condciz})-(\ref{condcivp})
are defined on a maximal interval of existence
$(t_{\rm{min}},t_{\rm{max}})$ with $t_{\rm{min}}<0$ and $t_{\rm{max}} \leq t^+$, which
depends a priori on the initial condition and parameters (in particular, $\e$).
From (\ref{condciz})-(\ref{condcivp}), one can infer that
$\| X(0) \|_p < M\, \e$ (the size of $Q_{\rm{a}}$ in (\ref{estimqa}))
for $\e$ small enough. Let
\begin{equation}
\label{defteps}
t_\e = {\rm{sup}}\, \{ t\in (0,{\rm{min}}(T\, \e^{1-\alpha} , t_{\rm{max}} )): \
\forall s \in (0,t), \,
\| X(s) \|_p \leq M\, \e    \} .
\end{equation}
From (\ref{estimqa}) and the triangle inequality, we obtain
\begin{equation}
\label{estimarg}
\| Q_{\rm{a}}(t) + W(t)+R(t) \|_p \leq 2\, M\, \e ,  \ \ \  \forall \, t \in [0, t_\e ].
\end{equation}
In addition, from definition (\ref{defteps}) we have either
\begin{equation}
\label{case1}
t_\e < {\rm{min}}(T\, \e^{1-\alpha} , t_{\rm{max}} )
{\rm ~and~ } \| X(t_\e) \|_p = M\, \e
\end{equation}
or
\begin{equation}
\label{case2}
t_\e = T\, \e^{1-\alpha} < t_{\rm{max}}
\end{equation}
(if $\| X \|_p$ is bounded on $[0,t_{\rm{max}})$, then $t_{\rm{max}}=t^+ > T\, \e^{1-\alpha}$).
Integrating (\ref{eq:rbis}) twice yields
\begin{equation}
\label{duhamelr}
R(t)=\dot{r}(0)\, t + R(0) + \chi \sigma \int_0^t{ (t-s)\, G(Q_{\rm{a}} + W+R)(s)\, ds }.
\end{equation}
Using the fact that $t_\e \leq T\, \e^{1-\alpha}$, estimate (\ref{estimarg}) and
the first bound in (\ref{sizeG}), one obtains from the above identity the following inequality:
\begin{equation}
\label{estimr1}
\forall t \in [0, t_\e ], \
\| R(t) \|_p \leq T\, \e^{1-\alpha} \| \dot{R}(0) \|_p + \| {R}(0) \|_p +
C_1\, \sigma \, \e^{2 - \alpha}.
\end{equation}
Then the assumption (\ref{sizemu}) and
the property (\ref{condcivp}) yield
\begin{equation}
\label{estimr}
\forall t \in [0, t_\e ], \ \| R(t) \|_p \leq C\, \e^\alpha .
\end{equation}
Similarly, we have
\begin{equation}
\label{duhamelrp}
\dot{R}(t)=\dot{R}(0)  + \chi \sigma \int_0^t{  G(Q_{\rm{a}} + W+R)(s)\, ds },
\end{equation}
which implies
\begin{equation}
\label{estimrp}
\forall t \in [0, t_\e ], \ \| \dot{R}(t) \|_p \leq
\| \dot{R}(0) \|_p + O(\sigma \, \e)
\leq C\, \e^{2\alpha -1} .
\end{equation}
Moreover, using Duhamel's formula in (\ref{eq:W}) yields
$$
W(t)=\cos{t}\, W(0)+ \sin{t}\, \dot{W}(0) + \int_0^t{\sin{(t-s)}\, [N(W,R)](s)\, ds}.
$$
Recalling that $t_\e \leq T\, \e^{1-\alpha}$,
using (\ref{condciz}), and using
(\ref{estimqa}), (\ref{sizemu}),
(\ref{sizeG}) and (\ref{estimr}) to estimate $N(W,R)$ from the definition (\ref{defn}), we get
\begin{equation}
\label{estimz1}
\forall t \in [0, t_\e ], \
\| W(t) \|_p \leq C_1\, \e^{\alpha} + C_2\, \e^{\alpha-1}\,  \int_0^t{\| W(s)\|_p\, ds} .
\end{equation}
By Gronwall's lemma we then have
\begin{equation}
\label{estimz}
\forall t \in [0, t_\e ], \ \| W(t) \|_p \leq C_1\, e^{C_2\, T}\, \e^{\alpha}  .
\end{equation}
Similarly, we obtain
$$
\dot{W}(t)=-\sin{t}\, W(0)+ \cos{t}\, \dot{W}(0) + \int_0^t{\cos{(t-s)}\, [N(W,R)](s)\, ds}.
$$
Using the same estimates as the ones involved in proving (\ref{estimz1}) and estimate (\ref{estimz}),
one can show that the above identity yields
\begin{equation}
\label{estimzp}
\forall t \in [0, t_\e ], \ \| \dot{W}(t) \|_p \leq C\, \e^{\alpha}  .
\end{equation}
Summing estimates (\ref{estimr}), (\ref{estimrp}), (\ref{estimz}), (\ref{estimzp}), we find
that $\| X(t) \|_p = O(\e^{\alpha}) < M\, \e$ for all $t \in [0, t_\e ]$ if $\e$ is small enough.
Consequently, the property (\ref{case1}) is not satisfied, which implies that (\ref{case2}) must hold instead.
We therefore have
$$
t_\e = T\, \e^{1-\alpha}
$$
in estimates (\ref{estimr}), (\ref{estimrp}), (\ref{estimz}) and (\ref{estimzp}).
Combining these estimates with the DpS error bound (\ref{estimerrqa}) and the assumption (\ref{sizemu}),
we deduce the error bounds (\ref{estimu}), (\ref{estimv}) from the identities
$$
u(t)-2\e \, {\rm{Re}}\, (A(\e^{\alpha-1}t)\, e^{it})
=W(t)+R(t)+Q_{\rm{a}}(t)-2\e \, {\rm{Re}}\, (A(\e^{\alpha-1}t)\, e^{it}),
$$
$$
\dot{u}(t)+2\e \, {\rm{Im}}\, (A(\e^{\alpha-1}t)\, e^{it})
=\dot{W}(t)+\dot{R}(t)+\dot{Q}_{\rm{a}}(t)+2\e \, {\rm{Im}}\, (A(\e^{\alpha-1}t)\, e^{it}),
$$
$$
v = R - \sigma\, Q_{\rm{a}} - \sigma\, W.
$$
This completes the proof of Theorem \ref{th:dps}.

%%%%%%%%%%%%%%%%%%%%%%%%% sec: Time-periodic traveling waves %%%%%%%%%%%%%%%
\section{Time-periodic traveling waves}\label{sec:tw}
We now use the DpS equations to investigate time-periodic traveling wave solutions of our system. In what follows, it will be convenient to use the strain formulation \eqref{eq:HertzS} and also rewrite the DpS equation \eqref{eq:S1new} in terms of strain-like variables
\beq\label{eq:S1newS}
2i\k\omega^3\partial_\tau \d^-A_n=\d^+[h_\alpha(\delta^-B_n,|\delta^-A_n|)\delta^-A_n] - \d^+[h_\alpha(\delta^-B_{n-1},|\delta^-A_{n-1}|)\delta^-A_{n-1}].
\eeq
A special class of solution of \eqref{eq:S1newS}, \eqref{eq:S2new} has the form of a \emph{periodic traveling wave}
\bea\label{eq:An_tw}
\d^-A_n(\t) = a e^{i(\O\t - qn - \f)}, \quad \d^{-}B_{n}(\t)=b,
\eea
where the frequency $\O$ depends on the amplitude $a>0$, the constant $b$ and the wave number $q$ through the nonlinear dispersion relation
\bea\label{disp}
\O = \frac{2}{\k\o^3}\left|\sin \frac{q}{2}\right|^{2}h_{\a}(b, a).
\eea
Note that definition of $h_{\a}$ in \eqref{eq:aux2} requires that $b \le 2\k a$. In particular, the dispersion relation has a closed form when $b = 0$. In fact, as shown in \cite{GJ11} for $\k=1$, in this case
$$
h_\a(0,r) = 2^{\a-1}c_\a\k^\a r^{\a-1}, \quad c_\a = \frac{1}{\pi}\frac{\G(1/2)\G(\a/2+1)}{\G((\a+1)/2+1)},
$$
and the dispersion relation is given by
\beqs
\O_0 = \frac{c_\a 2^{\a}(\k a)^{\a-1}}{\o^3} \left|\sin \frac{q}{2}\right|^{2}.
\eeqs
Equations \eqref{eq:U0V0}, \eqref{eq:An_tw} and \eqref{disp} yield the following first-order approximation of the periodic traveling wave solutions of system \eqref{eq:HertzS}:
\beq\label{eq:xytw_app}
x_n^{tw}(t) =\e b + 2\k\e a \cos (nq - \o_{tw}t + \f), \q y_n^{tw}(t)  = \e b - \frac{2\k\e a}{\rho}\cos (nq - \o_{tw}t + \f),
\eeq
where $\o_{tw}$ is the traveling wave frequency given by
\bea\label{eq:w_tw}
\o_{tw} = \O\e^{\a -1} + \o = \o + \frac{2\e^{\a -1}}{\k\o^3}\left|\sin \frac{q}{2}\right|^{2}h_{\a}(b, a).
\eea

To investigate how well the dynamics governed by the DpS equations approximates the traveling wave solutions of \eqref{eq:HertzS}, we consider initial conditions determined from the first-order approximation \eqref{eq:U0V0}:
\beq
\begin{split}
x_n^{app}(t) =\e \d^-B_n(\e^{\a-1}t) + \k\e \d^-A_{n}(\e^{\a-1}t)e^{i\o t} + c.c.\\
y_n^{app}(t)  =\e \d^-B_n(\e^{\a-1}t) -\frac{\k\e}{\r}\d^-A_{n}(\e^{\a-1}t)e^{i\o t} + c.c.
\end{split}
\label{eq:xy_app}
\eeq
at $t = 0$, along with initial velocities given by
\beq
\begin{split}
\dot x_n(0) = \e\{\e^{\a-1}\d^-\dot B_n(0) + \k\e^{\a-1}\d^-\dot A_n(0) + i\k\o\d^-A_n(0) + c.c\}\\
\dot y_n(0) = \e\{\e^{\a-1}\d^-\dot B_n(0) - \frac{\k}{\r}\e^{\a-1}\d^-\dot A_n(0) - i\frac{\k}{\r}\o\d^-A_n(0) + c.c\}.
\end{split}
\label{eq:xy_v_app}
\eeq
Here $\d^-A_n(0)$ are set to be small perturbations of \eqref{eq:An_tw} at $\t = 0$ and $\f = 0$:
\beq\label{eq:dA0_ptb}
\d^-A_n(0) =a[ (1 + \zeta_n^{(1)})\cos(nq) - i(1 + \zeta^{(2)}_n)\sin(nq)],
\eeq
where $\zeta^{(1)}_n$ and $\zeta^{(2)}_n$ are uniformly distributed random variables in $[-\zeta, \zeta]$ with small $\zeta>0$. Let $C = g_\a(b, |a|) >0$ for given constants $a$ and $b < 2 \kappa |a|$.
Then $\d^-B_n(\t)$ is determined from \eqref{eq:S2new} by numerically solving
\beq\label{eq:g_alpha}
g_\alpha(\delta^-B_n(\t),|\delta^-A_n(\t)|)=C.
\eeq
From the definition of $g_\a$ in (\ref{eq:aux1}),
it is clear that the function $b \mapsto g_\a(b, r) $ is decreasing on $(-\infty , 2 \kappa r )$
and satisfies $\lim_{b\rightarrow -\infty}{g_\a(b, r)}=+\infty$ and $g_\a(b, r)=0$ for $b \geq 2 \kappa r$.
Consequently equation (\ref{eq:g_alpha}) admits a unique solution
$\delta^-B_n(\t) \in (-\infty , 2 \kappa |\delta^-A_n(\t)|)$.
In particular, we obtain the value of $\d^-B_n(0)$ by solving (\ref{eq:g_alpha}) at $\t = 0$.
Note that $\d^-\dot A_n(0)$ and $\d^-\dot B_n(0)$ can be computed exactly using DpS equations \eqref{eq:S1newS}, \eqref{eq:g_alpha}, although their contribution is negligible since they correspond to higher order terms in the expression of initial velocities $\dot x_n(0)$ and $\dot y_n(0)$.
In particular, when the perturbation is zero (i.e. $\zeta = 0$), the initial condition simplifies to
\beq
\begin{split}
&x_n(0) = \e b + 2\e\k a\cos(nq + \f), \q y_n(0) = \e b - \frac{2\e\k a}{\r}\cos(nq + \f), \\
&\dot x_n(0) = 2\o_{tw}\k\e a\sin(nq + \f), \q \dot y_n(0) = -\frac{2\o_{tw}\k\e a}{\r}\sin(nq + \f).
\end{split}
\label{eq:ini_tw}
\eeq
Integrating \eqref{eq:HertzS} numerically on a finite chain with these initial conditions and using periodic boundary conditions $x_0 = x_{N}$, $x_{N+1} = x_{1}$, we can compare the solution of \eqref{eq:HertzS} (referred to in what follows as the numerical solution) with the ansatz \eqref{eq:xytw_app} when $\zeta = 0$ or the ansatz \eqref{eq:xy_app} when $\zeta > 0$. The latter is obtained by solving  \eqref{eq:S1newS}, \eqref{eq:g_alpha} with
periodic boundary conditions $\d^-A_0(\t) = \d^-A_N(\t)$, $\d^-A_{N+1}(\t) = \d^-A_{1}(\t)$, using the Runge-Kutta method and a standard numerical root finding routine to determine $\d^-B_n(\t)$ from \eqref{eq:g_alpha} for given $\d^-A_n(\t)$ at each step.

\subsection{Numerical traveling waves and the DpS approximation}
We now consider a locally resonant chain with $N = 50$ masses. In what follows, we set $\k = 1$, noting that other values of this parameter can be recovered by the appropriate rescaling of time and amplitude. To investigate the accuracy of the DpS approximation of the small-amplitude traveling waves, we first fix mass ratio $\r=1/3$ and normalize the traveling wave solution \eqref{eq:An_tw} by fixing $a =1$, $\f = 0$ and $b =1$. The linear frequency is given by $\o = \sqrt{\k + \k/\r} = 2$.

\begin{figure}[htp]
\centering
\centerline{\psfig{figure=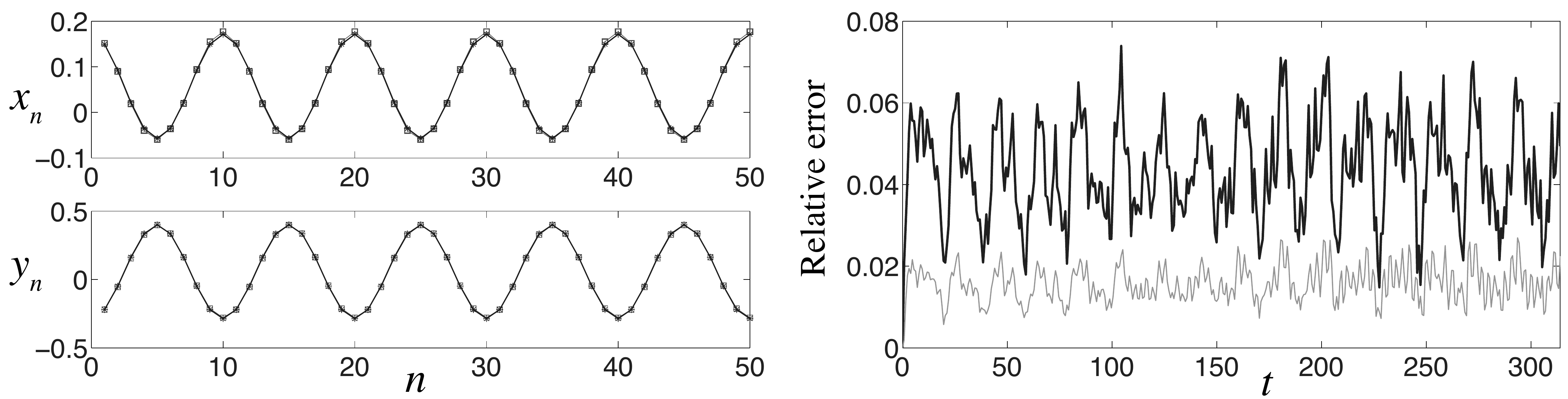,width=\textwidth}}
\caption{\footnotesize Left plot: strain profiles of small-amplitude approximate solution \eqref{eq:xytw_app} (connected stars) and numerical solution of \eqref{eq:HertzS} (connected squares) at $t = 100\, T_{tw} \approx 314$. Right plot: the relative errors $E_x(t)$ (black curve) and $E_y(t)$ (grey curve) of the DpS approximation. Here $\f = 0$, $k = 1$, $q = \pi/5$, $b = 1$, $a=1$, $\zeta=0$ and $\o_{tw} = w + 0.001$.}
\label{fig:tw_r03_q15pi}
\end{figure}

\begin{figure}[htp]
\centering
\centerline{\psfig{figure=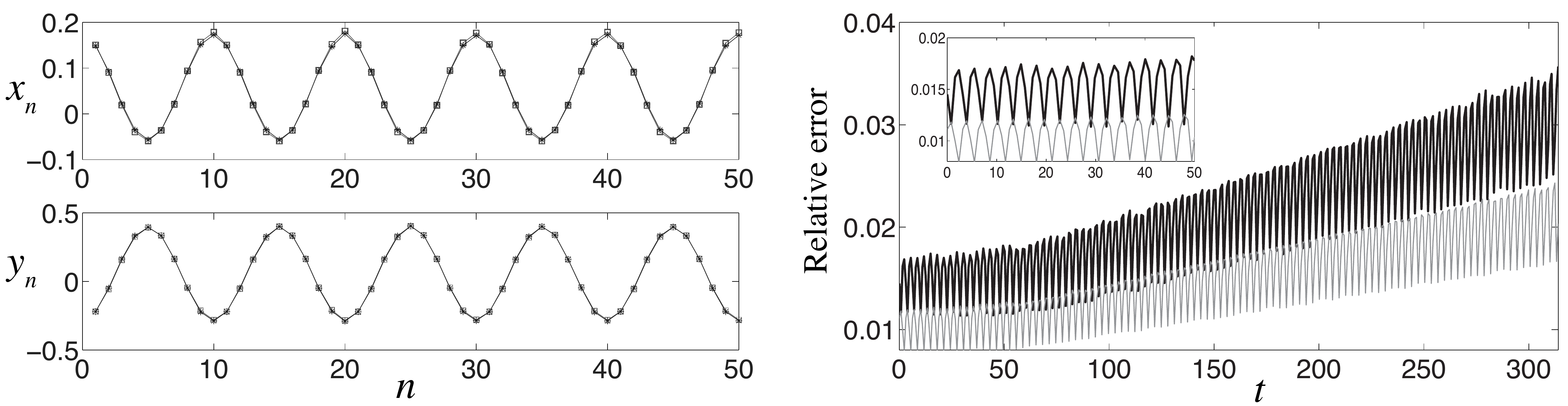,width=\textwidth}}
\caption{\footnotesize Left plot: strain profiles of small-amplitude approximate solution (connected stars) from the ansatz \eqref{eq:xy_app} and perturbed traveling wave solution (connected squares) at $t = 100\, T_{tw} \approx 314$. Right plot: the normalized differences $\wt E_x(t)$ (black curve) and $\wt E_y(t)$ (grey curve) of the perturbed traveling wave and the unperturbed solution shown in Fig.~\ref{fig:tw_r03_q15pi}. Here $\zeta = 0.01$ and all the other parameters are the same as in Fig.~\ref{fig:tw_r03_q15pi}. A growth
of the perturbations can be clearly observed in the dynamics.}
\label{fig:tw_r03_q15pi_ptd}
\end{figure}

In the first numerical run, we set $q=\pi /5$, $\zeta = 0$ and consider the traveling wave with frequency $\o_{tw} = \o + 0.001$, so that $\e \approx 0.057$ from the nonlinear dispersion relation \eqref{eq:w_tw}. Equation \eqref{eq:xytw_app} then yields the amplitude of $x_n^{tw}$ approximately equal to $0.114$, which corresponds to the small-amplitude regime. As shown in the left panel of Fig.~\ref{fig:tw_r03_q15pi}, the agreement between the numerical and approximate solutions is excellent, even after a long time $t = 100\, T_{tw}$, where $T_{tw} = 2\pi/\o_{tw} = 3.126$ is the period of the traveling wave. The relative errors of the approximate solutions
\beq\label{relerr_tw}
E_x(t) = \frac{1}{2\e\k a}||x^{app}_n(t) -x_n(t)||_{\infty} \q \text{and} \q E_y(t) = \frac{\r}{2\e\k a}||y^{app}_n(t) -y_n(t)||_{\infty}
\eeq
are less than $8\%$ at the final time of computation and remain bounded throughout the reported time evolution, as shown in the right panel of Fig.~\ref{fig:tw_r03_q15pi}.
In the second numerical run, we set $\zeta = 0.01$ for the perturbation in \eqref{eq:dA0_ptb}, while the other parameters remain the same. The agreement between the numerical and approximate solutions is still excellent over the same time interval (see the left plot of Fig.~\ref{fig:tw_r03_q15pi_ptd}). Moreover, the right plot of Fig.~\ref{fig:tw_r03_q15pi_ptd} shows the normalized differences
\beq\label{relerr_twPtd}
\wt E_x(t) = \frac{1}{2\e\k a}||x^{ptd}_n(t) -x_n(t)||_{\infty} \q \text{and} \q \wt E_y(t) = \frac{\r}{2\e\k a}||y^{ptd}_n(t) -y_n(t)||_{\infty}
\eeq
between a perturbed traveling wave $(x^{ptd}_n,y^{ptd}_n)$
(numerical solution obtained for the perturbed initial condition)
and the unperturbed numerical traveling wave solution $(x_n,y_n)$ shown in Fig.~\ref{fig:tw_r03_q15pi} with the wave number $q=\pi /5$. As shown by
Fig.~\ref{fig:tw_r03_q15pi_ptd}, the initial perturbation is not amplified at the early stage of the numerical integration of \eqref{eq:HertzS}
(for $t \leq 50 \approx 16\, T_{tw}$). However, the subsequent growth of perturbations indicates the
instability of the traveling wave solution.

\begin{figure}[htp]
\centering
\centerline{\psfig{figure=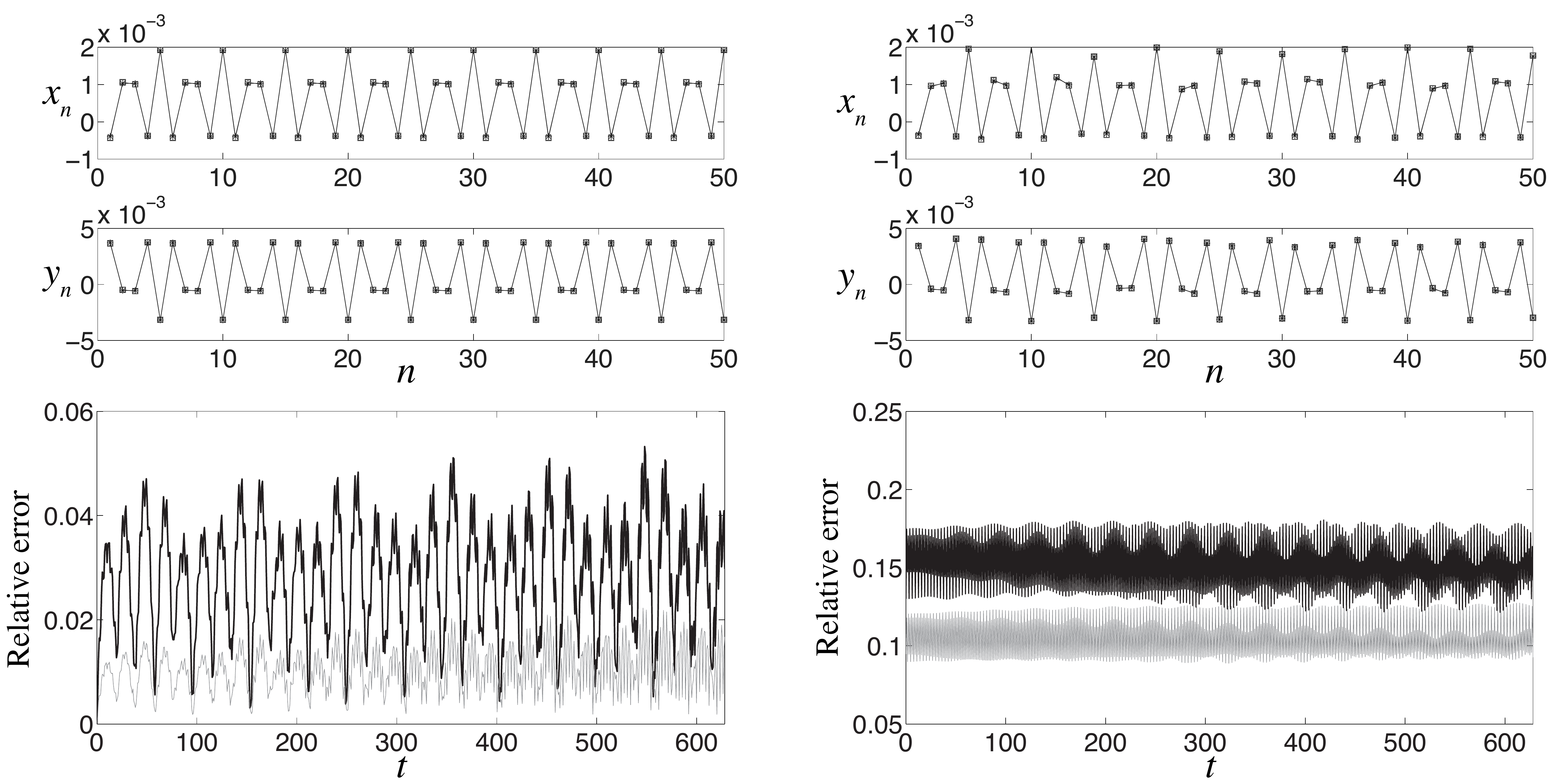,width=\textwidth}}
\caption{\footnotesize Top panels: strain profiles of small-amplitude approximate solution (connected stars) from the ansatz \eqref{eq:xy_app} and numerical
results (connected squares) with both unperturbed (left) and perturbed initial conditions with $\zeta = 0.1$ (right) at $t = 200\, T_{tw} \approx 628$, respectively. Bottom panels: left plot shows the relative errors $E_x(t)$ (black curve) and $E_y(t)$ (grey curve) of the DpS approximation. Right plot shows the normalized differences $\wt E_x(t)$ (black curve) and $\wt E_y(t)$ (grey curve) between the perturbed and unperturbed traveling waves. Here $q = 4\pi/5$ and all the other parameters are the same as in Fig.~\ref{fig:tw_r03_q15pi}.}
\label{fig:tw_r03_q45pi}
\end{figure}

In the next computation, we increase the wave number up to $q = 4\pi/5$ and keep all the other parameters the same as before. Now the asymptotic scale becomes quite small as $\e \approx 6.37 \times 10^{-4}$, which yields a very small amplitude of $x_n^{tw} \approx 0.0013$.
As shown in Fig.~\ref{fig:tw_r03_q45pi}, the DpS equations can successfully capture the dynamics of the small-amplitude traveling wave solution of the original system. In addition,
the small-amplitude traveling wave with $q = 4\pi/5$ appears to be stable on the interval $t \in [0,600]$.
However, this result may be linked with the very small traveling wave amplitude,
and instabilities might appear on longer time scales.

\begin{figure}[htp]
\centering
\centerline{\psfig{figure=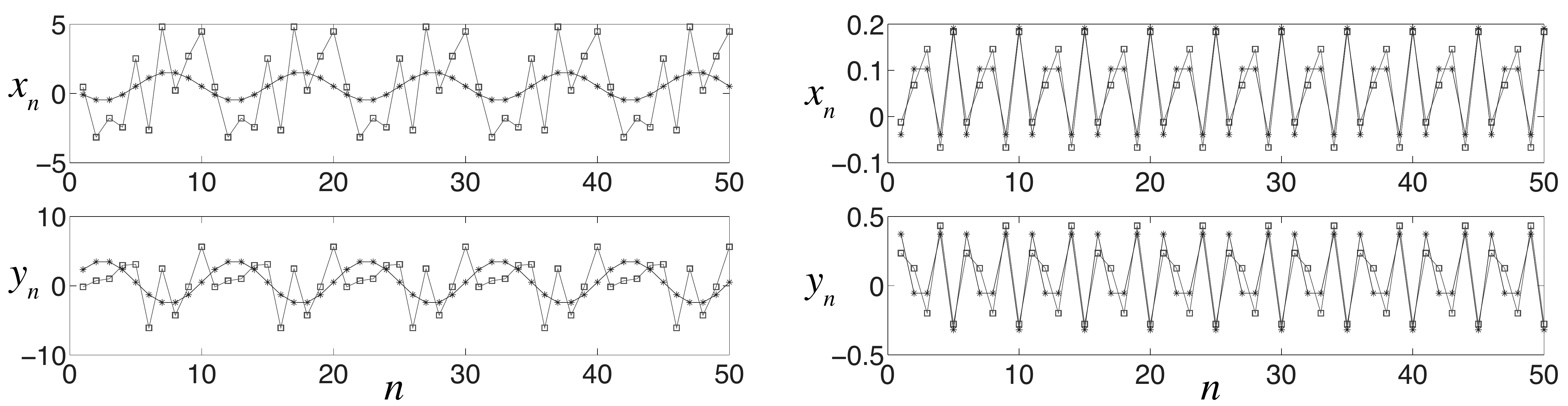,width=\textwidth}}
\caption{\footnotesize Left panel: results of the simulations with the same parameters as in Fig.~\ref{fig:tw_r03_q15pi} except for $\o_{tw} = \o + 0.003$ and the snapshot is taken at $t = 10\, T_{tw} \approx 31.4$. Right panel: results of the simulations with the same parameters as in the left panels of Fig.~\ref{fig:tw_r03_q45pi} except for $\o_{tw} = \o + 0.009$ and the snapshot is taken at $t = 314$.}
\label{fig:tw_r03_breakdown}
\end{figure}

It is worth pointing out that the time scale of the validity of the DpS approximation depends not only on the asymptotic scale $\e$ but also on the wave number $q$. To illustrate this, we first consider the wave number $q = \pi/5$ but increase the traveling wave frequency up to $\o_{tw} = \o + 0.003$, which yields $\e \approx 0.514$ and the amplitude of $x_n^{tw}$ around $1.028$. As revealed by the left plot in Fig.~\ref{fig:tw_r03_breakdown}, the DpS equations fail to describe the dynamics of \eqref{eq:HertzS} appropriately soon after we start the integration. We then increase the wave number up to $q = 4\pi/5$ but choose $\o_{tw} = \o + 0.009$ yielding $\e \approx 0.0516$, which is even smaller than the asymptotic scale $\e$ used in Fig.~\ref{fig:tw_r03_q15pi}. However, notable difference between the traveling wave patterns of the numerical and approximate solutions are observed over the same time interval $[0, 314]$ (see the right plot of Fig.~\ref{fig:tw_r03_breakdown}).

\subsection{Effect of mass ratio $\r$}
\begin{figure}[htp]
\centering
\centerline{\psfig{figure=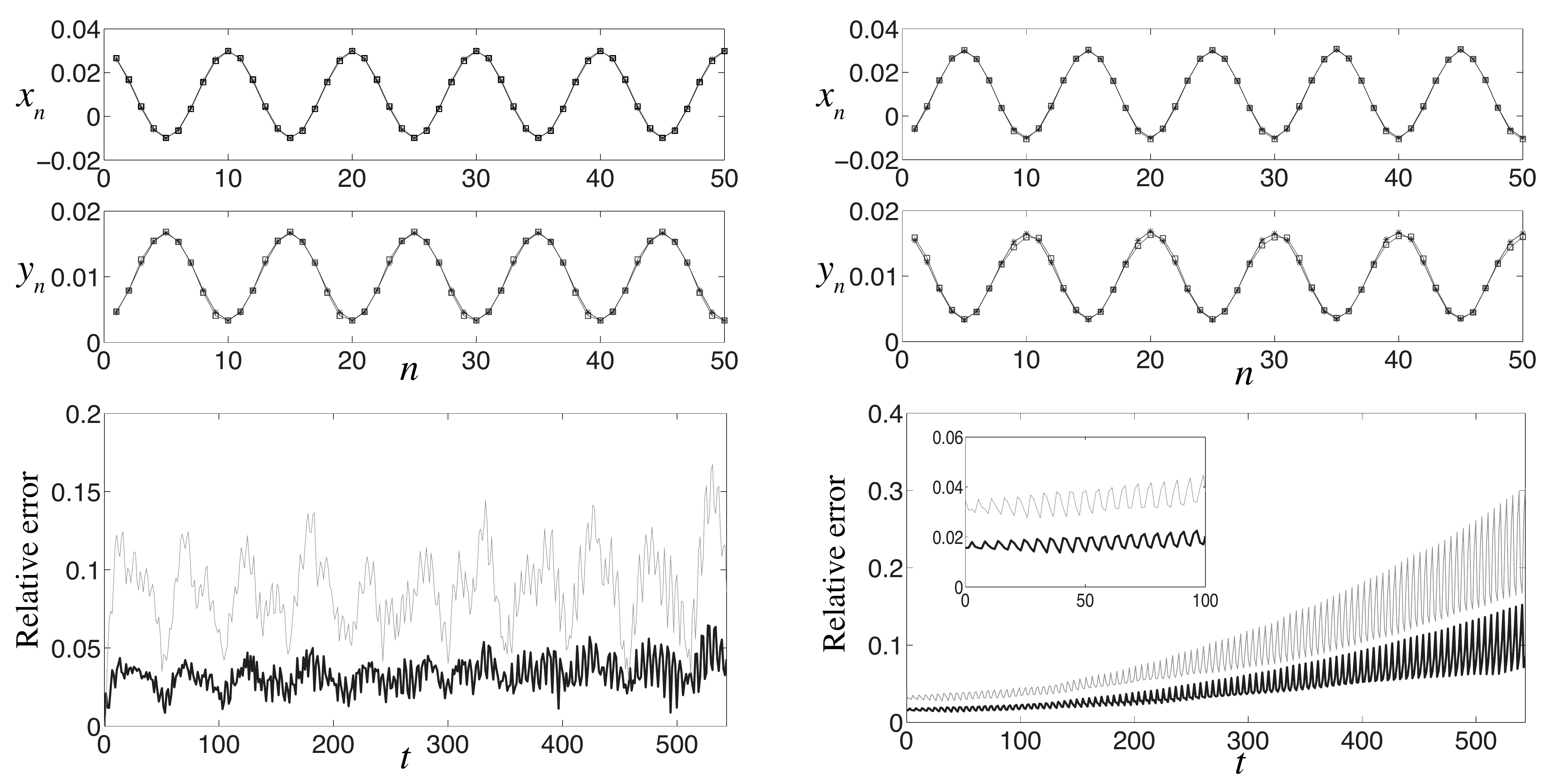,width=\textwidth}}
\caption{\footnotesize Results of the simulations with the same parameters as in Fig.~\ref{fig:tw_r03_q45pi} except for
$\rho =3$, $q = \pi/5$ and $\e = 0.01$. The strain profiles of both panels
correspond to snapshots taken at $t = 100\, T_{tw}$ and the perturbation added in the initial condition is $\zeta = 0.01$.}
\label{fig:tw_r3_q15pi}
\end{figure}

To investigate the effect of the mass ratio $\r$ on the validity of the DpS approximation of the small-amplitude traveling waves, we fix $q = \pi/5$, $\e = 0.01$ and keep the other parameters the same as before. We choose mass ratio $\r = 3$ and the traveling wave frequency in \eqref{eq:w_tw} is now given by $\o_{tw} \approx \o + 0.002$, where the linear frequency is $\o = 1.1547$. The results of the simulations are shown in Fig.~\ref{fig:tw_r3_q15pi}. We observe again that the agreement between the numerical and approximate solutions remains excellent over the time interval $[0, 100\, T_{tw}]$, and the
numerical traveling waves appear to be stable over the time interval [0, 100]. However, we note the growing trend of the difference between the perturbed and unperturbed traveling wave solutions after $t \approx 100$, which illustrates the instability of the traveling wave.

\begin{figure}[htp]
\centering
\centerline{\psfig{figure=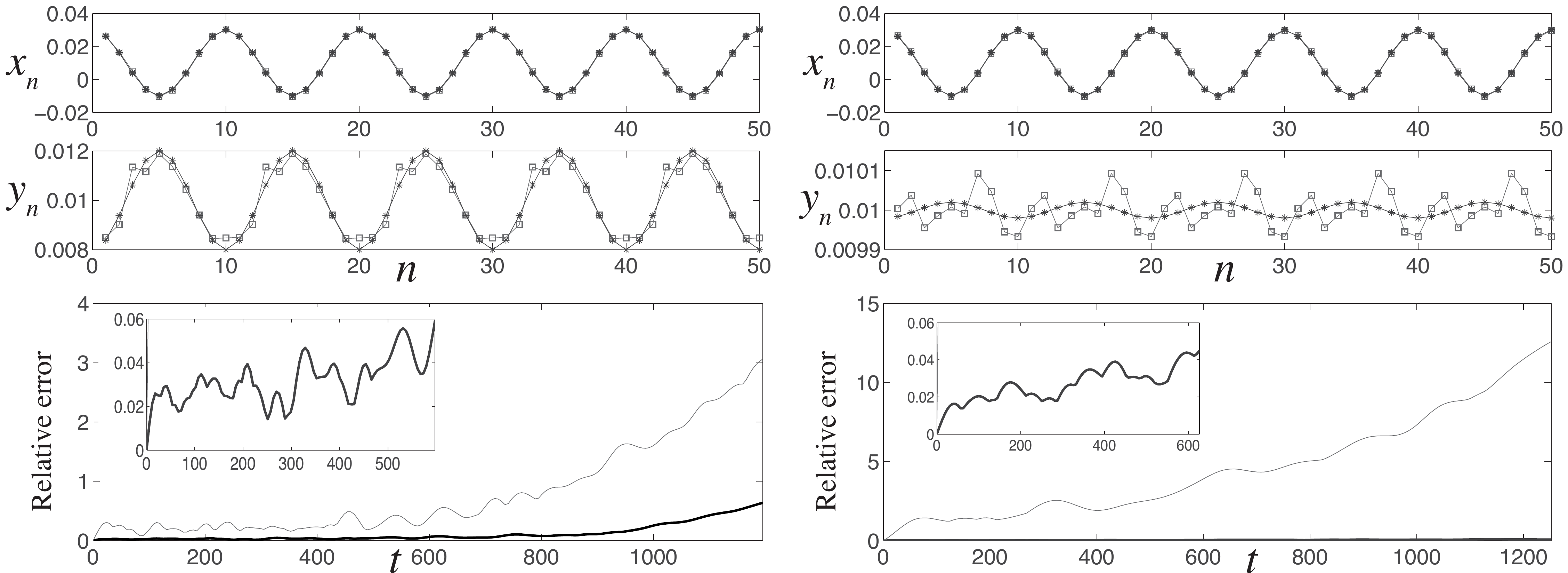,width=\textwidth}}
\caption{\footnotesize Results of the simulations with the same parameters as in the left panel of Fig.~\ref{fig:tw_r3_q15pi} except for $\rho = 10$ (left panel) and $\rho = 1000$ (right panel). The strain profiles of both panels
correspond to the snapshot taken at $t = 100\, T_{tw}$. The insets in the bottom plots show the enlarged plots of the relative error $E_x(t)$.}
\label{fig:tw_r10r1000}
\end{figure}

We further increase $\r$ to the critical value $\r_c =\e^{1-\a} =10$. The linear and traveling wave frequencies are given by $\o = 1.0488$ and $\o_{tw} = 1.0517$, respectively. As shown by the left panel of Fig.~\ref{fig:tw_r10r1000}, the wave form of $y_n$ is no longer sinusoidal at $t = 100\, T_{tw}$, while the wave form of $x_n$ remains sinusoidal and matches its DpS approximation very well over the time interval $[0, 100\, T_{tw}]$. We observe also a growing trend in the relative error between the numerical and approximate
solutions, despite the structural similarities of the profiles.
To investigate the case when $\r$ is large ($O(\e^{1-\g}), \g > \a$), we now set $\r = 1000$ and keep all other parameters the same as before. The linear frequency is now given by $\o = 1.0005$ and approximate frequency is $\o_{tw} = 1.0038$. Again, over the integration time interval $[0, 100\, T_{tw}]$ the wave form of $x_n$ remains sinusoidal and close to the ansatz \eqref{eq:xytw_app} but the waveform of $y_n$ is non-sinusoidal and significantly deviates from the DpS traveling wave approximation (see the right plots in Fig.~\ref{fig:tw_r10r1000}). These results are consistent with the discussion in Sec.~\ref{sec:derivation}, since the derivation of the DpS equations \eqref{eq:S1new}-\eqref{eq:S2new} requires that $\rho$ is below the critical value $\r_c = \e^{1-\a}$ for given $\e$. Note that if we further set $b = 0$, the DpS equation \eqref{eq:S1new} reduces to the one capturing the traveling waves in Newton's cradle problem in \cite{GJ11}. Numerical results (not reported here) reveal that the numerical solution of $x_n$ is sinusoidal and very well approximated by the traveling wave ansatz \eqref{eq:xytw_app}. However, the difference between exact and approximate solutions of $y_n$ is very substantial. Here the structural characteristics of the solution for $y_n$ are no longer properly captured by the DpS approximation.

These numerical simulations reveal that the validity of the DpS approximation at fixed wave number $q$ is very sensitive to the mass ratio $\r$. When $\r$ is relatively small, the generalized DpS equations \eqref{eq:S1new}-\eqref{eq:S2new} successfully capture the dynamics of periodic traveling waves. When $\r \ge \r_c = \e^{1-\a}$, an increasing deviation between the exact and approximate solutions of $y_n$ emerges. However, in all cases, the agreement of approximate and numerical solutions for $x_n$ remains excellent over a finite time interval. In addition, traveling wave instabilities can be observed
depending on the values of $\rho$, wave number $q$, wave amplitude and time scales considered.

%%%%%%%%%%%%%%%%%%%%% sec: bright breathers %%%%%%%%%%%%%%%%%%%%%%%%%%%%%%%%%%%%%%%
\section{Bright Breathers}\label{sec:br}

Bright breathers are time-periodic solutions of (\ref{eq:Hertz}) which converge to constants (zero strain) at infinity, i.e.
\beq\label{po:vanish}
u_n , v_n \rightarrow c_{\pm}, \q \mbox{as} \q n \rightarrow \pm\infty
\eeq
uniformly in time. In this section we examine the existence of either exact or {\em long-lived} bright breather solutions of (\ref{eq:Hertz}).
The second class of solutions refers to
spatially localized solutions of (\ref{eq:Hertz}) remaining close to a time-periodic
oscillation over long times.

We begin by noting the existence of trivial exact bright breather solutions of (\ref{eq:Hertz}) for which particles do not interact,
i.e. $(u_{n-1} (t) - u_{n} (t))_{+}^\a =0$ for all $t$ and $n$. This is equivalent to having
\begin{equation}
\label{constraint}
u_{n-1} (t) \leq u_{n} (t)\ \ \  \forall\, t\in \mathbb{R}, \, \forall n \in \mathbb{Z},
\end{equation}
\beq
\begin{split}
\ddot u_{n} = \k(v_n - u_n), \\
\r\ddot v_{n}= \k(u_{n} - v_{n}).
\end{split}
\label{eq:linear}
\eeq
The time-periodic solutions of (\ref{eq:linear}) read
\begin{equation}
\label{uncpl}
u_n (t)=\frac{\rho}{1+\rho}\, a_n\, \cos{(\omega\, t + \phi_n)} + b_n, \ \ \
v_n (t) = - \frac{1}{1+\rho}\, a_n\, \cos{(\omega\, t + \phi_n)} + b_n,
\end{equation}
where we can fix $a_n \geq 0$ and denote
$\omega^2 = \kappa\, (1+1/\rho )$.
Bright breather profiles are obtained for
\begin{equation}
\label{cv}
\lim_{n \rightarrow \pm \infty}a_n =0, \ \ \
\lim_{n \rightarrow \pm \infty}b_n = c_\pm .
\end{equation}
A solution of (\ref{eq:Hertz}) is obtained if and only if the constraint (\ref{constraint}) is satisfied, which is equivalent to
\begin{equation}
\label{constraint2}
b_n - b_{n-1} \geq \frac{\rho}{1+\rho}\, [\,  a_n^2 + a_{n-1}^2 -2 a_n a_{n-1} \cos{(\phi_n - \phi_{n-1} )} \, ]^{1/2}
\ \ \ \forall \, n\in \mathbb{Z}.
\end{equation}
For $(a_n )_{n \in \mathbb{Z}} \in \ell_1 (\mathbb{Z})$, this is equivalent to assuming
$$
b_n = d_n +
\frac{\rho}{1+\rho}\, \sum_{k=-\infty}^n{[\, a_k^2 + a_{k-1}^2 -2 a_k a_{k-1} \cos{(\phi_k - \phi_{k-1} )}  \, ]^{1/2}},
$$
where $(d_n )_{n \in \mathbb{Z}}$ is a nondecreasing sequence converging as $n \rightarrow \pm \infty$.
It is clear from this expression that $(b_n )_{n \in \mathbb{Z}}$ corresponds to a kink profile. Moreover,
fixing $\phi_n - \phi_{n-1} = \pi$, we can simplify the above expression to obtain
$$
b_n =
d_n + \frac{\rho}{1+\rho}\, \big(a_n + 2 \sum_{k=-\infty}^{n-1}{a_k   }\big).
$$
We now prove the following.

\begin{thm}
\label{nobreath}
The trivial bright breather solutions defined by (\ref{uncpl})-(\ref{cv})-(\ref{constraint2}) are the only time-periodic bright breather solutions of (\ref{eq:Hertz}).
\end{thm}

To prove Theorem \ref{nobreath}, we follow the method of the proof of nonexistence of breathers in FPU chains with repulsive interactions given in \cite{JKC13}. Suppose $(u_n, v_n)$ is a bright breather, i.e. a $T$-periodic solution of \eqref{eq:Hertz} satisfying \eqref{po:vanish}. Adding the equations in \eqref{eq:Hertz}, one can see that the bright breather solution must satisfy
\beq\label{eq:uv_ddot}
\ddot u_{n} + \r\ddot v_{n} =(u_{n-1} - u_n )_{+}^{\a} - (u_n - u_{n+1})_{+}^{\a} .
\eeq
Integrating \eqref{eq:uv_ddot} over one period, we obtain
\beq
\bar F_{n+1} = \bar F_{n} , \qq \bar F_n = \frac{1}{T}\int_0^T (u_{n-1}(t) - u_n(t))_{+}^\a dt  .
\eeq
Note that $\bar F = \bar F_n$ is independent of $n$ and one can show that it vanishes. Indeed, by \eqref{po:vanish}, we have
\beq
\lim_{n \rightarrow \pm \infty} ||u_n  - u_{n-1}||_{L^{\infty}(0,T)} = 0.
\eeq
Meanwhile,
\beq\label{eq:F}
|\bar F| = \frac{1}{T}\int_0^T (u_{n-1}(t) - u_{n}(t))_{+}^\a dt  \le ||u_{n-1}(t) - u_{n}(t) ||_{L^{\infty}(0,T)}^\alpha
\eeq
for all $n$. Taking the limit $n \rightarrow \pm\infty$ in (\ref{eq:F}) one obtains $\bar F = 0$. Consequently, for each $n$, we have
\beq
 \int_0^T (u_{n-1}(t) - u_{n}(t))_{+}^{\a} dt = 0
\eeq
and since $F_n = (u_{n-1} - u_{n})_{+}^\a$ is non-negative, continuous and $T$-periodic, we have $F_n = 0$ for all $t$ and $n$. Thus $(u_n, v_n)$ satisfies the linear system (\ref{eq:linear}) and corresponds to a trivial bright breather solution.
This completes the proof of Theorem \ref{nobreath}.\\

%\vspace{1ex}

In what follows we show that, although nontrivial time-periodic bright breathers do not exist for system
(\ref{eq:Hertz}), long-lived small-amplitude bright
breather solutions can be found when $\rho$ is large. This is due to the connection between
(\ref{eq:Hertz}) and the DpS equation (\ref{eq:DpSlr}) established in section \ref{sec:derivationrholarge}.
Equation (\ref{eq:DpSlr}) admits time-periodic solutions of the form
\begin{equation}
\label{solper}
A_n (\tau ) = \varepsilon\,  a_n\, e^{i\, \omega_0 \, |\varepsilon|^{\alpha -1}\, \tau},
\end{equation}
where $a=(a_n)_{n \in \mathbb{Z}}$ is a real sequence and $\varepsilon\in \mathbb{R}$
an arbitrary constant, if and only if $a$ satisfies
\begin{equation}
\label{dpsstat}
a= - \delta^+[|\delta^-a|^{\alpha -1}\delta^-a].
\end{equation}
In particular, nontrivial solutions of (\ref{dpsstat})
satisfying $\lim_{n\rightarrow \pm\infty}{a_n}=0$
correspond to bright breather solutions of (\ref{eq:DpSlr})
given by (\ref{solper}).
These solutions are doubly exponentially decaying,
so that they belong to $\ell_p$ for all $p\in[1,\infty]$.
They have been studied
in a number of works (see \cite{JS14} and references therein).
The following existence theorem
for spatially symmetric breathers has been proved in \cite{JS14}
using a reformulation of (\ref{dpsstat}) as a two-dimensional mapping.

\begin{thm}
\label{homoclinic}
The stationary DpS equation (\ref{dpsstat}) admits
solutions $a_n^i$ ($i=1,2$) satisfying
$$
\lim_{n\rightarrow \pm\infty}a_n^i=0,
$$
$$
(-1)^n\, a_{n}^i >0, \quad
|a_n^i | > |a_{n-1}^i | \quad \text{for all } \, n\leq 0,
$$
$$
\text{and} \quad
a_n^1 =a_{-n}^1,  \ \ \
a_n^2 = -a_{-n+1}^2,
 \ \ \  \mbox{ for all } n \in \mathbb{Z}.
$$
Furthermore, for all $\beta \in(0,1)$, there exists $n_0\in\mathbb{N}$
such that the above-mentioned solutions $a_n^i$ satisfy,
for $i=1,2$:
$$
\forall n \geq n_0, \quad |a_n^i| \leq \beta^{1+\alpha^{n-n_0}} .
$$
\end{thm}

Considering the bright breather solutions of
(\ref{eq:DpSlr}) given by (\ref{solper}) with $a=a^i$, $\e=1$,
and applying Theorem \ref{th:dps}, one
obtains stable exact solutions of equations
(\ref{eq:Hertz}), close to the bright breathers, over the corresponding time scales.
This yields the following result formulated for $\kappa =1$.

\begin{thm}
\label{longlivedb}
Fix constants $C_{\rm{r}}, C_{\rm{i}},T>0$.
Consider a solution $a^i=(a_n^i)_{n\in \mathbb{Z}}$ ($i=1,2$) of
the stationary DpS equation (\ref{dpsstat}) described in Theorem
\ref{homoclinic}. There exist $\e_T , C_T >0$ such that for all
$\e \in (0, \e_T ]$
and for $\rho^{-1} \leq C_{\rm{r}}\, \e^{2(\alpha -1)}$,
for all initial condition of \eqref{eq:Hertz}
in $\ell_p^4$ satisfying
\begin{equation}
\label{condic}
\| u(0)-2\e \, a^i \|_p + \| \dot{u}(0) \|_p + \| v(0) \|_p
\leq C_{\rm{i}}\e^{\alpha}, \ \ \
\| \dot{v}(0) \|_p \leq C_{\rm{i}}\e^{2\alpha -1},
\end{equation}
the solution to equation \eqref{eq:Hertz}
is defined at least for $t\in[0,T\e^{1-\alpha}]$ and satisfies
\begin{equation}
\label{erreurb}
\| u(t)-2\e \, a^i\, \cos{(\Omega_b\, t)} \|_p
+ \| \dot{u}(t)+2\e \,  a^i\, \sin{(\Omega_b\, t)} \|_p \leq C_{T}\e^{\alpha},
\textrm{ for all }t\in[0,T\e^{1-\alpha}],
\end{equation}
with $\Omega_b = 1+\omega_0\, \e^{\alpha -1}$,
$\omega_0$ defined in (\ref{eq:defom0})
and
\begin{equation}
\label{estimv2}
\| v(t) \|_p \leq C_{T}\e^{\alpha}, \ \ \
\| \dot{v}(t) \|_p \leq C_{T}\e^{2\alpha -1},
\textrm{ for all }t\in[0,T\e^{1-\alpha}].
\end{equation}
\end{thm}

It is important to stress the differences between
the long-lived bright breather solutions provided by Theorem \ref{longlivedb}
and the trivial bright breathers analyzed at the beginning of this section.
The oscillations described in Theorem \ref{longlivedb} are nontrivial in the sense that
Hertzian interactions do not vanish identically. In addition, they are truly localized (for $p<\infty $)
whereas the trivial exact breathers are superposed on a nonvanishing kink component $b_n$.
Moreover, the (approximate) frequency $\Omega_b$ of long-lived bright breathers
satisfies $0 < \Omega_b - 1 =  O(\e^{\alpha -1})$ for breathers
with amplitude $O(\e )$.
For trivial exact bright breathers, the frequency $\omega$ is independent of amplitude and satisfies $0<\omega -1 = O(1/\rho)$
when $\rho$ is large. Under the assumptions of Theorem \ref{longlivedb}, we have
$1/\rho = O(\e^{2(\alpha -1)})$ and thus
$\Omega_b > \omega$ for $\e$ small enough.

\begin{figure}[htp]
\centering
\centerline{\psfig{figure=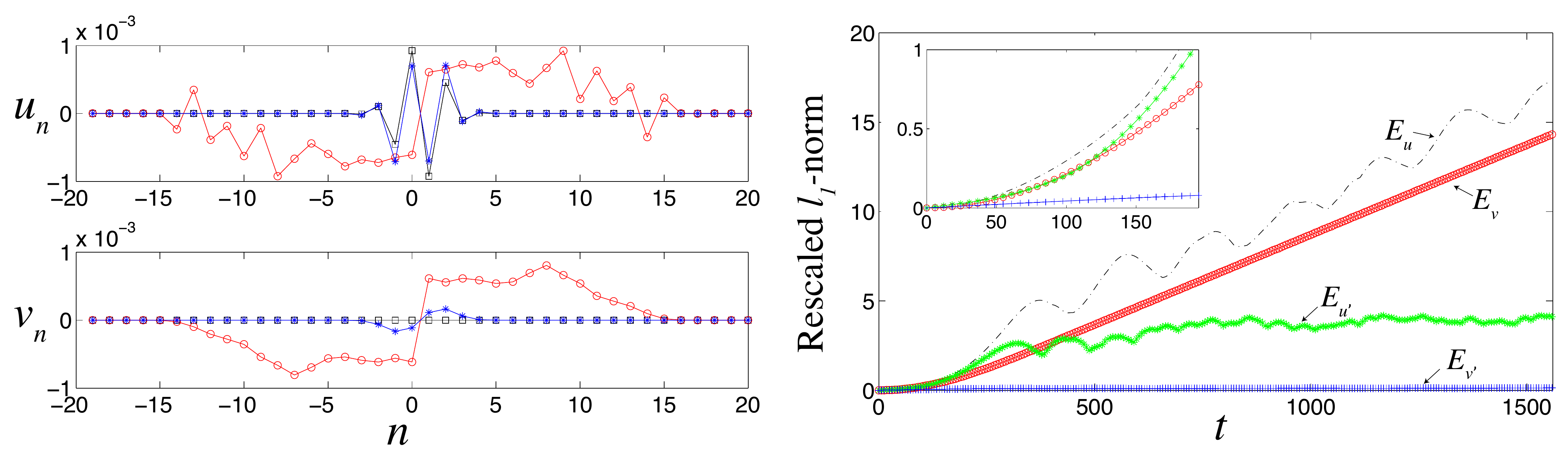,width=\textwidth}}
\caption{\footnotesize Left panels: snapshot of numerical solution of \eqref{eq:Hertz} at $t = 30\, T_b < 10\, T$ (blue stars) and $t = 80\, T$ (red circles), respectively, starting with initial conditions \eqref{ini_num} with $i=2$ (black squares). Right panel: time evolution of the rescaled $\ell_1$-norms defined in \eqref{diff_u_l1norm} and \eqref{diff_v_l1norm}, where $E_{u}(t)$ is represented by connected dots (black curve), $E_{\dot u}(t)$ by connected stars (green curve), $E_{v}(t)$ by connected circles (red curve) and $E_{\dot v}(t)$ by connected pluses (blue curve). The inset in the right plot shows time evolution of the same $\ell_1$-norms when $t \le T\e^{1-\a}$.}
\label{fig:long_lived_bb}
\end{figure}

 We now investigate the behavior of long-lived bright breathers numerically. Fixing $T = 2\pi/\o_0 = 19.4158$, $C_r = C_i = 1$ and choosing $\e = 0.01$, we consider a locally resonant chain of $N = 40$ masses with the mass ratio $\rho = 1000$ so that the inequality $\r^{-1} \le C_{r}\e^{2(\a-1)}$ in Theorem \ref{longlivedb}
 is satisfied. We then integrate the system \eqref{eq:Hertz} over a long time interval $[0, 80\, T]$, starting with initial conditions
\beq\label{ini_num}
u(0) = 2\e a^i , \quad v(0) = 0, \quad \dot u(0) = \dot v(0) = 0
\eeq
where $a^i$ is the numerical solution of \eqref{dpsstat} obtained using the method in \cite{JKC13}. The numerical simulation yields spatially localized solutions of \eqref{eq:Hertz} that stay close to the time-periodic oscillation
\beq\label{time-periodic_osc}
\wt u(t) = 2\e a^i \cos(\O_b t), \q \dot{\tilde{u}}(t) = -2\e a^i \sin(\O_b t) , \q \wt v(t) = \dot{\tilde{v}}(t) \equiv 0
\eeq
with $i=2$ over the time interval $[0, T\e^{1 - \a}]=[0,194.158]$, as can be seen in the inset of the right panel of Fig~\ref{fig:long_lived_bb}. Note that the comparison is made at times corresponding to multiples of $T_b = 2\pi/\O_b = 6.0862$. To measure the relative difference of the numerical solution of \eqref{eq:Hertz} and the time-periodic oscillations \eqref{time-periodic_osc}, we define the rescaled $\ell_1$-norms as follows:
\beq \label{diff_u_l1norm}
E_{u}(t) = \frac{||u(t) - \wt u(t)||_1}{\e^\a}, \q E_{\dot u}(t) = \frac{||\dot u(t) - \dot{\tilde{u}}(t)||_1}{\e^\a}
\eeq
and
\beq\label{diff_v_l1norm}
E_{v}(t) = \frac{||v(t) - \wt v(t)||_1}{\e^\a}, \q E_{\dot v}(t) =\frac{||\dot v(t) - \dot{\tilde{v}}(t)||_1}{\e^{2\a-1}}.
\eeq
The fact that those rescaled norms remain small over time interval $[0, T\e^{1 - \a}]$ is consistent with the result of Theorem~\ref{longlivedb} and confirms the existence of the long-lived bright breathers. However, at larger time, part of the energy is radiated away from the vicinity of the initially excited sites. As a result, we observe the breakdown of the localized structure for a long-time ($t \gg T\e^{1-\alpha}$) evolution, and the solution profile spreads out and eventually approaches a kink-type structure shown by circles in the left panel of Fig~\ref{fig:long_lived_bb}. This is associated with the growing magnitude of $v_n$ during the simulation.
%

%%%%%%%%%%% sec: Approximate dark breather solutions %%%%%%%%%%%%%%%%%%%%%%%%%
\section{Approximate dark breather solutions}\label{sec:db_approx}
We now turn to dark breather solutions, which, as we will see, are
fundamentally different from the waveforms considered in the
previous section and are not excluded by the results of Theorem~\ref{nobreath}.
To construct approximate dark breather solutions, we
start by
considering \emph{standing wave} solutions of the generalized DpS equations \eqref{eq:S1new}-\eqref{eq:S2new} in the form
\beq
\label{eq:ab_sw}
\d^-A_n(\t) = \d^-a_ne^{i(\O\t + \f)}
\ \
( a_n \in \hR ),
\ \ \
\d^{-}B_n(\t) = \d^-b_n,
\eeq
where $a_n$ and $b_n$ are time-independent.
Introducing $\o_{b} = \o + \O\e^{\a -1}$, we find that the first-order approximate solution of \eqref{eq:Hertz} reads
\beq
\begin{split}
u_n^{sw}(t) = \e b_n + 2\k\e a_n\cos(\o_{b}t + \f),\\
v_n^{sw}(t) = \e b_n - \frac{2\k}{\r}\e a_n\cos(\o_{b}t + \f).
\end{split}
\label{eq:uv_sw}
\eeq
Substituting \eqref{eq:ab_sw} in the generalized DpS equations \eqref{eq:S1new}-\eqref{eq:S2new} we obtain
\beq
\begin{split}
&-\mu a_n = h_{\a}(\d^-b_{n+1}, |\d^-a_{n+1}|)\d^-a_{n+1} -  h_{\a}(\d^-b_{n}, |\d^-a_{n}|)\d^-a_{n}\\
&\d^+ g_{\a}(\d^-b_n, |\d^-a_n|)=0,
\end{split}
\label{eq:DpSsw}
\eeq
where $\mu = 2\k\o^{3}\O = 2\k\o^{3}(\o_{b} - \o)\e^{1-\a}$.

Following the approach in \cite{GJ11}, for $\m \neq 0$ one can further show that $\wt a_n = |\m|^{\frac{1}{1 - \a}} a_n$, $ \wt b_n = |\m|^{\frac{1}{1 - \a}} b_n$ satisfy the renormalized equation
\beq
\begin{split}
&-\mbox{sign}(\mu) \wt a_n = h_{\a}( \d^-\wt b_{n+1}, |\d^-\wt a_{n+1}|) \d^-\wt a_{n+1} - h_{\a}(\d^-\wt b_{n}, |\d^-\wt a_{n}|)\d^-\wt a_{n} \\
&\d^+ g_{\a}(\d^-\wt b_n, |\d^-\wt a_n|)=0.
\end{split}
\label{eq:DpSsw1}
\eeq
where $\mbox{sign}(\m) = 1$ for $\m >0$ and $\mbox{sign}(\m) = -1$ for $\m < 0$. For simplicity we drop the tilde in \eqref{eq:DpSsw1} in what follows. %Observe that multiplying the first equation in \eqref{eq:DpSsw1} by $a_n$ and summing over $n$, one obtains that for all $\{a_n\} \in l_2(\hZ)$
%\beqs
%\mbox{sign}(\m)\sum_{n\in \hZ}a^2_n = \sum_{n\in\hZ}h_{\a}( \d^- b_{n+1}, |\d^- a_{n+1}|) |\d^- a_{n+1}|^2
%\eeqs
%where $h_{\a}(b, r)$ is nonnegative, which implies that a nontrivial solution for $\{a_n\}$ can be found only if $\m > 0$.
Numerical results suggest that a nontrivial solution for $\{\d^-a_n\}$ can be found if and only if $\mu > 0$. Thus it suffices to consider the case $\mu = 1$. It is convenient to rewrite \eqref{eq:DpSsw1} in terms of $\d^-a_n$ and $\d^-b_n$ by subtracting from the first equation in \eqref{eq:DpSsw1} the same equation at $n-1$:
\beq
\begin{split}
& -\d^-a_n = h_{\a}(\d^-b_{n+1}, |\d^-a_{n+1}|)\d^-a_{n+1} - 2 h_{\a}(\d^-b_{n}, |\d^-a_{n}|)\d^-a_{n} + h_{\a}(\d^-b_{n-1}, |\d^-a_{n-1}|)\d^-a_{n-1}\\
&g_{\a}(\d^-b_{n+1},|\d^-a_{n+1}|)=g_{\a}(\d^-b_n, |\d^-a_n|).
\end{split}
\label{eq:DpSsw2}
\eeq

\begin{figure}[htp]
\centering
\centerline{\psfig{figure=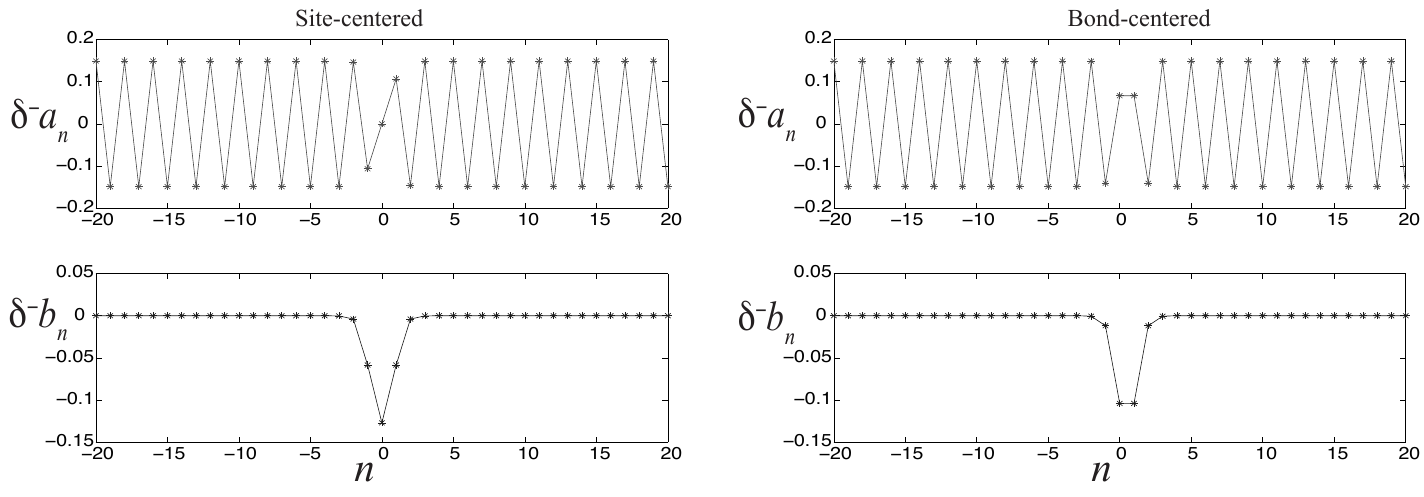,width=\textwidth}}
\caption{\footnotesize Left panel: A site-centered solution of \eqref{eq:DpSsw2} with $\d^-b_{-N} = c = 0$. Right panel: bond-centered solution.}
\label{fig:sw_anbn}
\end{figure}

We now use Newton's iteration to solve \eqref{eq:DpSsw2} numerically for $\d^-a_n$, $\d^-b_n$, $n=-N,\dots,N$, with periodic boundary conditions ($\d^-a_{-N-1} = \d^-a_N$ and $\d^-a_{N+1} = \d^-a_{-N}$). Note that the associated Jacobian matrix is singular due to the structure of second equation in \eqref{eq:DpSsw2}, and therefore an additional constraint is necessary. It is sufficient to fix $\d^-b_{-N} = c$, where $c$ is a constant.
Since we are looking for dark breathers, it is natural to consider initial values of $\d^-a_n$ in the form
\beq\label{eq:an_ini}
\d^-a^0_n = (-1)^n\mbox{tanh}(n - n_0)
\eeq
where $n_0$ is an arbitrary constant corresponding to spatial translation. One can then use the second equation in \eqref{eq:DpSsw2} to solve for initial guess of $\d^-b_n^0$, $n = -N+1, ..., N$. A standard Newton's iteration procedure of the system \eqref{eq:DpSsw2} with $4N + 1$ variables $\d^-b_{-N+1},\dots,\d^-b_N$, $\d^-a_{-N},\dots,\d^-a_N$ is then performed with the tolerance of $10^{-8}$.
%{\bf How does the DpS iteration then proceed ? With $b_n$ fixed an
%iteration for $a_n$ takes place, or subsequently (more likely)
%Eqs. (81) are concurrently solved for $a_n$ and $b_n$. Some
%clarification on the numerical procedure (and the tolerance used) would
%be useful here.}
Setting $n_0 = 0$ in \eqref{eq:an_ini} results in a site-centered solution shown in the left panel of Fig.~\ref{fig:sw_anbn}, whereas the bond-centered solution corresponds to $n_0 = 1/2$ shown the right panel. Typical breather waveforms of both bright~\cite{Theo10} and
dark~\cite{Chong13} type come in these two broad families~\cite{Flach08}.

Once the Newton's iteration converges to a fixed point $\{(\d^-a^*_n, \d^-b^*_n)\}$, we can compute the first-order approximate solution of \eqref{eq:HertzS} given by
\beq
\begin{split}
x_n^{sw}(t) = |\bar\m|^{\frac{1}{\a-1}} \{\d^-b^*_n + 2\k \d^-a^*_n\cos(\o_{b}t + \f)\},\\
y_n^{sw}(t) = |\bar\m|^{\frac{1}{\a-1}} \{\d^-b^*_n - \frac{2\k}{\r} \d^-a^*_n\cos(\o_{b}t + \f)\},
\end{split}
\label{eq:xy_sw}
\eeq
where $\bar\m = 2\k\o^{3}(\o_{b} - \o)$.
%To compute the Jacobian matrix in each step of the iteration, one needs to numerically evaluate the following expressions:
%\beas
%&\frac{\partial}{\partial \d^-b_n} g_{\a}(\d^-b_n, |\d^-a_n|) =  - \frac{\a}{2\pi}\int_{0}^{2\pi}(- \d^-b_n + 2\k| \d^-a_n|\cos t)_+^{\a-1} dt\\
%&\frac{\partial}{\partial \d^-a_n}g_{\a}( \d^-b_n, |\d^-a_n|) =  \frac{\d^-a_n}{|\d^-a_n|} \cd \frac{\a\k}{\pi}\int_{0}^{2\pi}(- \d^-b_n + 2\k| \d^-a_n|\cos t)_+^{\a-1} \cos tdt\\
%&\frac{\partial}{\partial \d^-b_n} \wt h_{\a}(\d^-b_n, |\d^-a_n|)=  - \frac{\a}{2\pi}\int_{0}^{2\pi}e^{-it}\frac{(- \d^-b_n + 2\k| \d^-a_n|\cos t)_+^{\a-1}}{|\d^-a_n|} dt\\
%&\frac{\partial }{\partial \d^-a_n}(\wt h_{\a}(\d^-b_n, | \d^-a_n|)\cd \d^-a_n) = \frac{\a\k}{\pi}\int_{0}^{2\pi}e^{-it}(-\d^-b_n + 2\k|\d^-a_n|\cos t)_+^{\a-1} \cos t dt
%\eeas
%Note that when $|\d^-a_n|\rightarrow 0$, we use the fact that
%\beas
%&\frac{\partial}{\partial \d^-a_n} g_{\a}(\d^-b_n, 0) = 0\cd 2\k^2\a(\a-1)(-\d^-b_n)^{\a-2} = 0\\
%&\frac{\partial}{\partial \d^-b_n} h_{\a}(\d^-b_n, 0) = \k\a(1-\a)(-\d^-b_n)^{\a-2}.
%\eeas
%
%
%
\section{Numerically exact dark breathers and linear stability analysis}\label{sec:db_exact}
Having constructed the initial seed \eqref{eq:xy_sw}, we can compute the numerically exact dark breather solution of system \eqref{eq:HertzS} with periodic boundary conditions. Let $x(t)$, $y(t)$, $\dot x(t)$ and $\dot y(t)$ denote the row vectors with component $x_n(t)$, $y_n(t)$, $\dot x_n(t)$ and $\dot y_n(t)$, respectively. Let $Z(t) = (x(t), y(t))$. We seek time-periodic solutions $(Z(t), \dot Z(t))$ of the Hamiltonian system \eqref{eq:HertzS} satisfying the initial conditions $(Z(0), \dot Z(0))$. For a fixed period of the dark breather solution given by $T_b = 2\pi/\o_b$, where $\o_b$ is the breather frequency, the problem is equivalent to finding the fixed points of the corresponding Poincar\'e map $P_{T_b}[ (Z(0) ,\dot Z(0))^{T} ] = ( Z(T_b), \dot Z(T_b) )^{T}$. Since the
system \eqref{eq:HertzS} has the time-reversal symmetry, we can further restrict the
solution space by setting $\dot Z(0) \equiv 0$.

We use a Newton-type algorithm (see, for example, Algorithm 2 in \cite{Aubry97}) to compute the fixed point.
More precisely, let $\D Z(0)$ be the small increment of the initial data that needs to be determined. It is then sufficient to minimize $||P_{T_b}[(Z(0)+\D Z(0), 0)^T] - (Z(0)+\D Z(0),0)^T||$ at each iteration step. Notice that $P_{T_b}[(W(0), 0)^T]$ can be approximated by $P_{T_b}[(Z(0), 0)^T] + M(T_b) (\D Z(0), 0)^T$
for sufficiently small $\D Z(0)$. Here $M(t)$ is the associated monodromy matrix of the variational equations satisfying
\beq\label{eq:var}
\frac{d}{dt} M(t) = \mathcal{J}(Z(t), \dot Z(t) ) M(t), \quad M(0)=I,
\eeq
where $\mathcal{J}(Z(t), \dot Z(t))$ is the Jacobian matrix of the nonlinear system \eqref{eq:HertzS} at $(Z(t), \dot Z(t))$, and $I$ is the identity matrix. The Jacobian for the Newton's iteration is then given by $I - M(T_b)$ and it is singular since it can be shown that $M(T_b)$ has an eigenvalue pair equal to 1. To remove the singularity, we impose the additional constraint that the time average of $x_1(t)+\D x_1(t)$, the first component $Z(t)+\D Z(t)$, is fixed to be (approximately) zero. Observing that $(Z(t)+\D Z(t), \dot Z(t)+\D \dot Z(t))^T \approx (Z(t),\dot Z(t))^T + M(t) (Z(0), 0)^T$, we obtain
\beq\label{eq:extra_constraint}
\frac{1}{T_b}\int_{0}^{T_b} x_1(t) dt + \frac{1}{T_b}\int_{0}^{T_b} M_1(t) \cd \D Z(0) dt = 0,
\eeq
where $M_1(t)$ is the first row of $M(t)$.

In the results discussed below we set $\k = 1$. To characterize the solution, we define the vertical centers of the solution for $x$ and $y$ components \cite{Chong13},
\beq\label{eq:center}
C_x = \frac{\sup_{t\in[0,T_b]}x_1(t) + \inf_{t\in[0,T_b]}x_1(t) }{2},  \q C_y = \frac{\sup_{t\in[0,T_b]}y_1(t) + \inf_{t\in[0,T_b]}y_1(t) }{2}
\eeq
and the amplitudes of the breather,
\beq\label{eq:amp}
K_x = \frac{ \sup_{t\in[0,T_b]}x_1(t) - \inf_{t\in[0,T_b]}x_1(t) }{2}, \q  K_y = \frac{ \sup_{t\in[0,T_b]}y_1(t) - \inf_{t\in[0,T_b]}y_1(t) }{2}.
\eeq
Note that the vertical center $C_x$ is approximately zero due to the constraint \eqref{eq:extra_constraint}. However, one can fix any other value of $C_x$ by replacing zero in the right hand side of \eqref{eq:extra_constraint} by $C_x$. To further investigate the long-term behavior of the dark breather solution, we introduce the relative error
\beq\label{eq:relerr}
E_b(t) = ||Z(mT_b)- Z(0)||_{\infty} / ||Z(0)||_{\infty}
\eeq
where $m = \lfloor t/T_b \rfloor $ and $Z(mT_b) = (x(mT_b), y(mT_b))$ represents the strain profile after integrating \eqref{eq:HertzS} over $m$ multiple of time periods, starting with the static dark breather $Z(0)$ as the initial condition.

\begin{figure}[htp]
\centering
\centerline{\psfig{figure=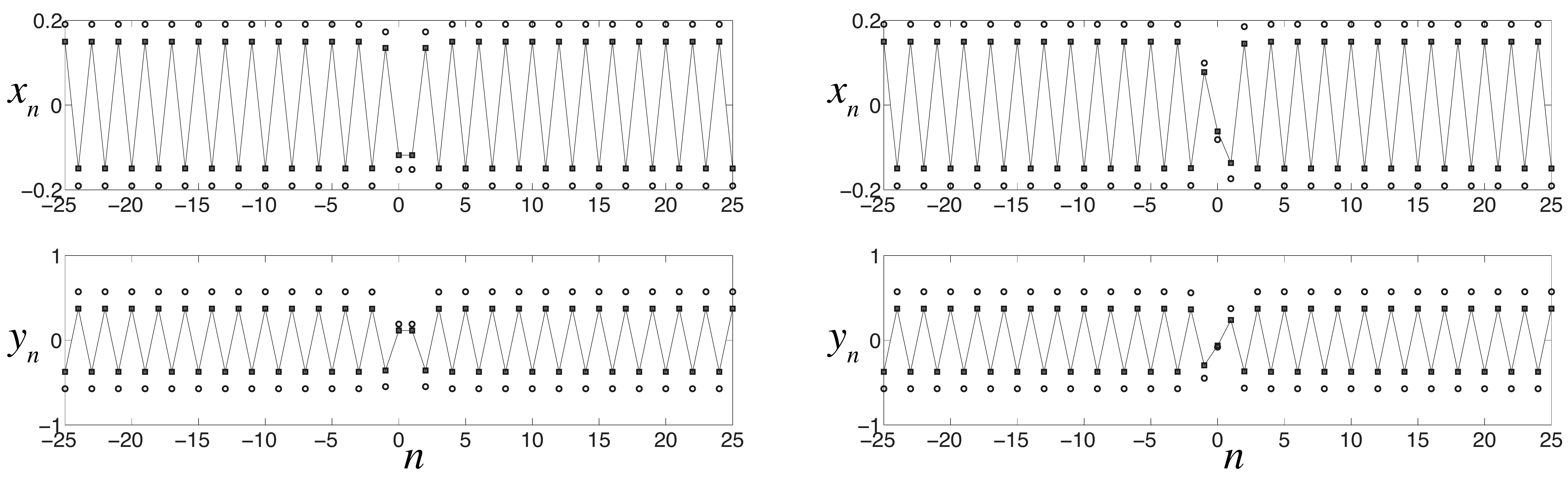,width=\textwidth}}
\caption{\footnotesize Left panel: a bond-centered solution of dark breather solution (connected stars) with frequency $\o_{b} = 2.05$. Connected squares represent strain profile after integration over $T_b$ and circles represent the DpS approximate solution from the ansatz \eqref{eq:xy_sw}. Right panel: site-centered solution. The relative errors $E_b(T_b)$ of both site-centered and bond-centered solutions are less than $4.5\times10^{-9}$. Here $\k =1$ and $\r = 1/3$.}
\label{fig:sw_xnyn_w205}
\end{figure}

We first consider the mass ratio $\r = 1/3$, so that the linear frequency is $\o = 2$. We choose a value of $\o_{b}$ that is slightly greater than this value but close enough to it in order to obtain a good initial seed with a small amplitude. Once the Newton-type solver converges to a numerically exact dark breather solution, we use the method of continuation to obtain an entire family of dark breathers that corresponds to different values of $\o_{b}$. Sample profiles of both bond-centered and site-centered dark breather solutions with $\o_{b} = 2.05$ along with the DpS approximate solutions \eqref{eq:xy_sw} are shown in Fig.~\ref{fig:sw_xnyn_w205}. The amplitudes $K_x$ and $K_y$ increase with $\o_{b}$, and the solution approaches zero as $\o_b \rightarrow \o$.

\begin{figure}[htp]
\centering
\centerline{\psfig{figure=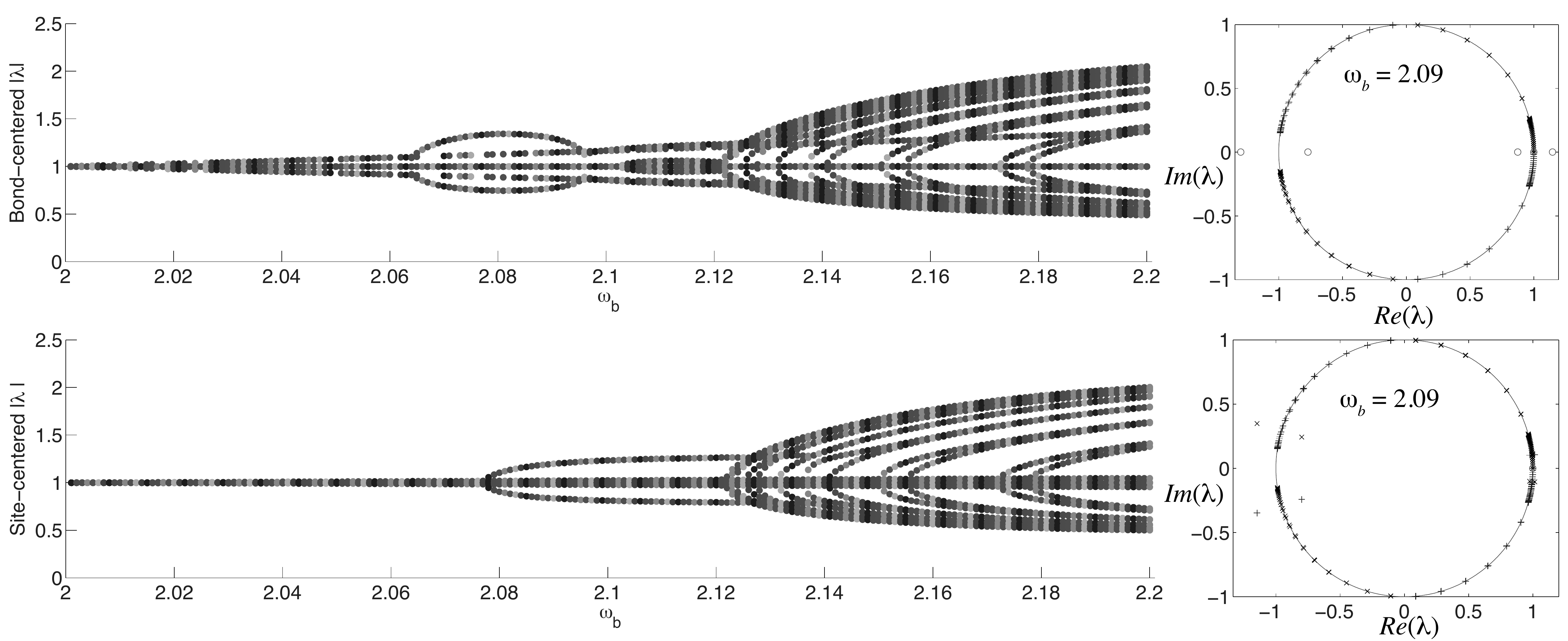,width=\textwidth}}
\caption{\footnotesize Left plots: moduli of the Floquet multipliers versus frequency $\o_{b}$ for the bond-centered (top) and site-centered (bottom) breathers. The Floquet multipliers for $\o_{b} = 2.09$ in the complex plane are shown in the respective plots on the right. Here $\k =1$ and $\r = 1/3$.}
\label{fig:r03floq}
\end{figure}

The linear stability of each obtained dark breather solution is examined via a standard Floquet analysis. The eigenvalues (Floquet multipliers) of the associated monodromy matrix $M(T_b)$ determine the linear stability of the breather solution. The moduli of Floquet multipliers for the site-centered and bond-centered solutions of various frequencies are shown in Fig.~\ref{fig:r03floq}, along with the numerically computed Floquet spectrum that
corresponds to the sample breather profile at $\o_{b}=2.09$. If any of these Floquet multipliers $\l_{i}$ satisfies $|\l_{i}|>1$, the corresponding breather is linearly unstable. We observed two types of instabilities in this Hamiltonian system. The first one is the \emph{real instability}, which corresponds to a real Floquet multiplier with magnitude greater than one; an example is shown in the right top plot in Fig.~\ref{fig:r03floq} for the bond-centered dark
breather with $\o_{b}=2.09$. The second type is the \emph{oscillatory instability}, which corresponds to a quartet of Floquet multipliers outside the unit circle with non-zero imaginary parts (see the right bottom plot of Fig.~\ref{fig:r03floq}
for the site-centered dark breather with $\o_{b}=2.09$).

\begin{figure}[htp]
\centering
\centerline{\psfig{figure=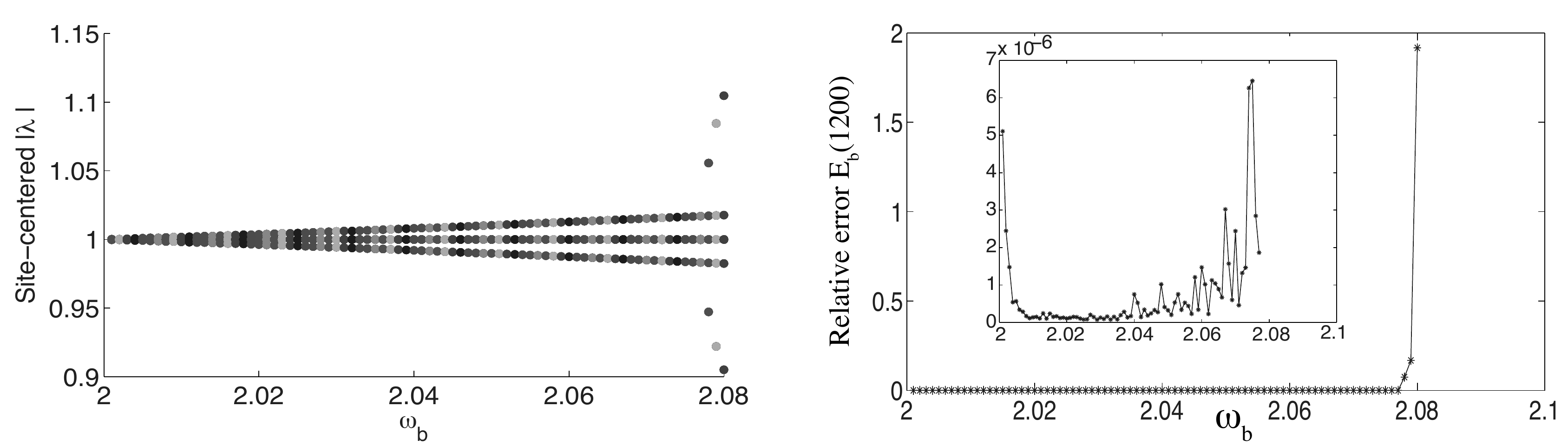,width=\textwidth}}
\caption{\footnotesize Left plot: moduli of the Floquet multipliers of the site-centered breathers for frequency $\o_{b} \le 2.08$. Right plot: the relative error $E_b(t)$ versus frequency at $t = 1200$. Inset shows the relative error for frequencies less than $\o_{b} \le 2.077$. Here $\k =1$ and $\r = 1/3$.}
\label{fig:r03sc_relerr}
\end{figure}

\begin{figure}[htp]
\centering
\centerline{\psfig{figure=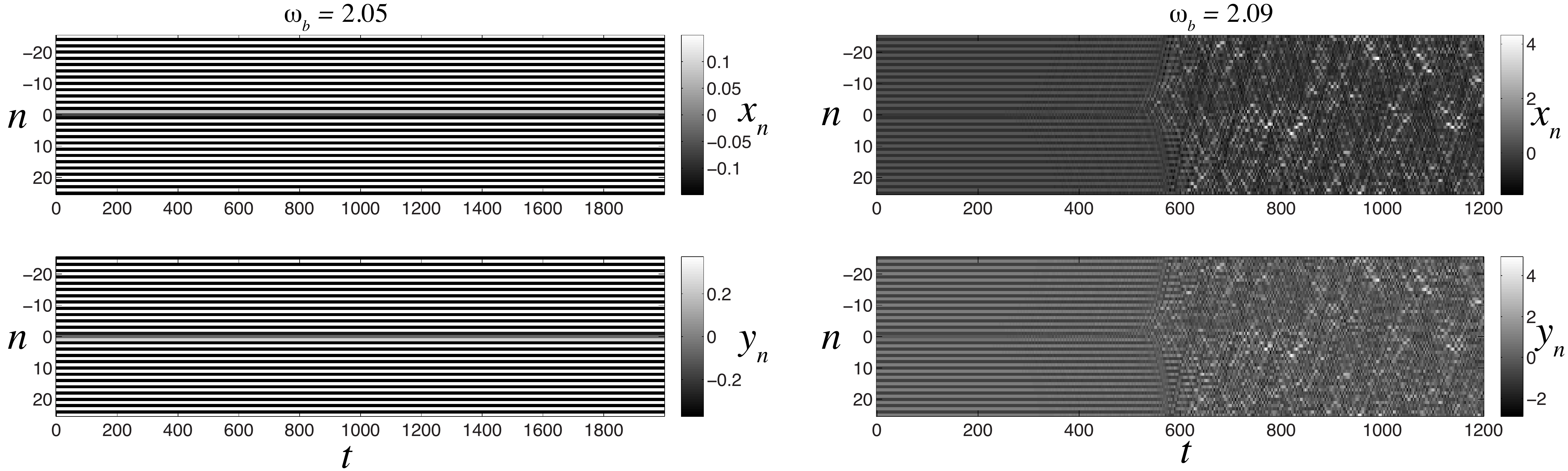,width=\textwidth}}
\caption{\footnotesize Left panel: contour plot of the time evolution of the site-centered solution for $\o_b = 2.05$. The color bar corresponds to the magnitude of the strain $x_n$ (top) and $y_n$ (bottom). Right panel: same computation as in left panel but for $\o_b = 2.09$. Here $\k =1$ and $\r = 1/3$.}
\label{fig:r03sc_evol}
\end{figure}

Numerical results reveal that the site-centered dark breathers appear to exhibit only oscillatory instabilities. These marginally unstable modes emerge at the beginning of the continuation procedure but remain weak until $\o_b$ reaches $\o_b \approx 2.078$. As shown in the right plot of Fig.~\ref{fig:r03sc_relerr}, the relative error $E_b(1200)$ stays below $7\times10^{-7}$ when $\o_b \le 2.077$ but increases dramatically afterwards. The results are consistent with the Floquet spectrum shown in the left plot of Fig.~\ref{fig:r03sc_relerr}. In fact, beyond the critical point of $\o_b \approx 2.078$, we observe the emergence of a new and stronger oscillatory instability mode that corresponds to two pairs of Floquet multipliers distributed symmetrically outside the unit circle around $-1$ (see also the right bottom plot of Fig.~\ref{fig:r03floq} for $\o_b = 2.09$). Representative space-time evolution diagrams of site-centered dark breather solutions are shown in Fig.~\ref{fig:r03sc_evol}, which suggests that the site-centered dark breather solutions with frequency close to the linear frequency $\o$ are long-lived and have marginal oscillatory instability, below
the pertinent critical point. However, the oscillatory instability becomes more and more significant as $\o_b$ increases, leading to the breakdown of the dark breather structure of the solution. In fact, beyond the critical
point, the breakup of the site-centered breather appears to
be accompanied in Fig.~\ref{fig:r03sc_evol} by a dramatic evolution,
whereby the configuration is completely destroyed and a form
of lattice dynamical turbulence ensues. This phenomenon is reminiscent of
traveling wave instabilities observed in \cite{james12} for Hertzian chains and
may be worth further study, which, however, is outside the scope of the
present manuscript.


\begin{figure}[htp]
\centering
\centerline{\psfig{figure=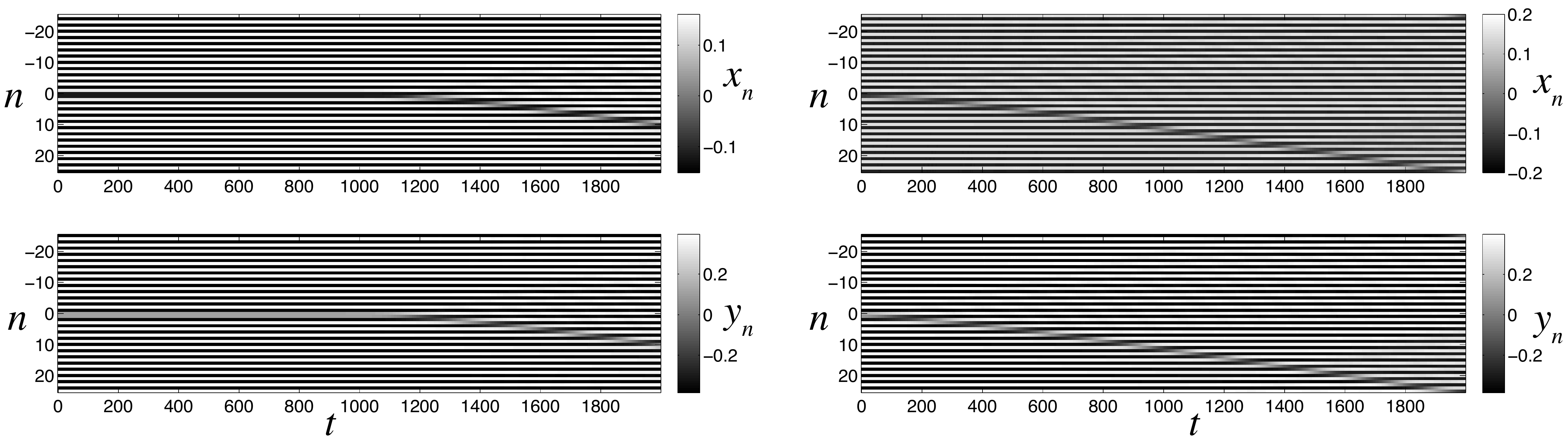,width=\textwidth}}
\caption{\footnotesize Left panel: contour plot of the time evolution of the bond-centered solution for $\o_b = 2.05$. The color bar corresponds to the magnitude of the strain $x_n$ (top) and $y_n$ (bottom). Right panel: same simulation as in the left panel but with the perturbed dark breather as the initial condition. Here $\k =1$ and $\r = 1/3$.}
\label{fig:r03bc_evol}
\end{figure}

\begin{figure}[htp]
\centering
\centerline{\psfig{figure=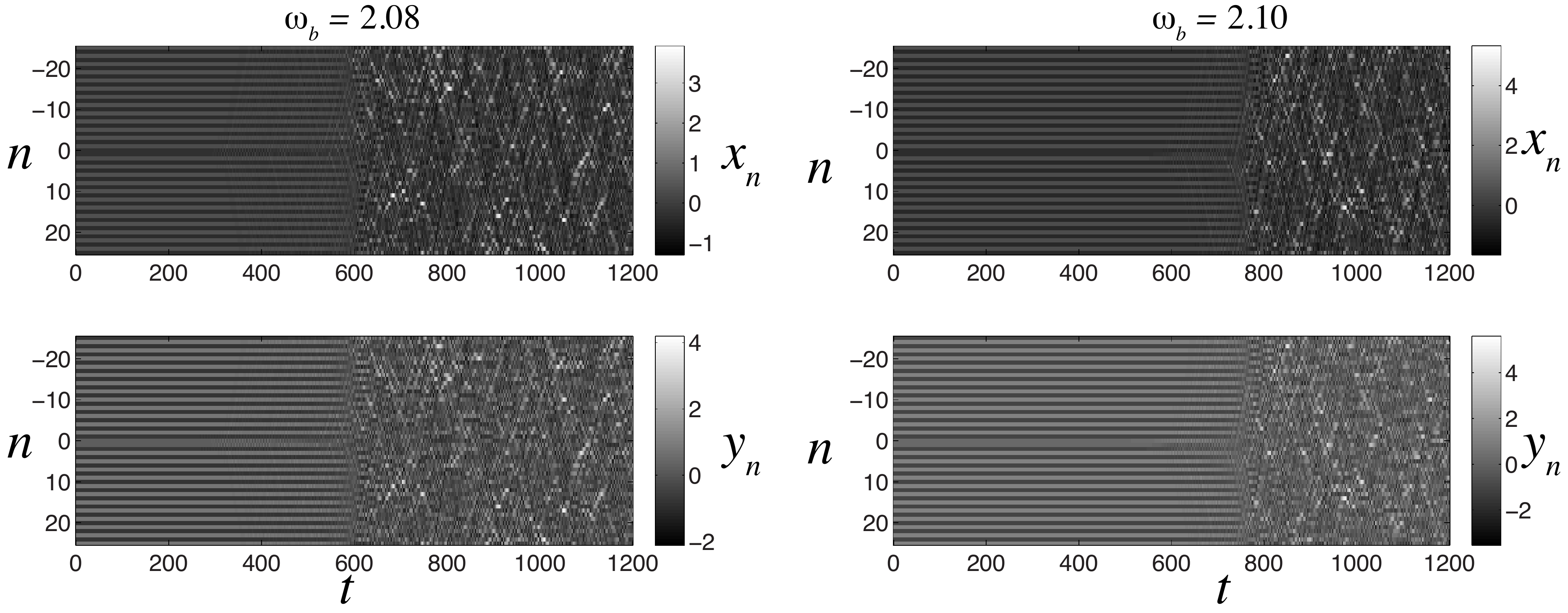,width=\textwidth}}
\caption{\footnotesize Left panel: contour plot of the time evolution of the bond-centered solution for $\o_b = 2.08$. The color bar corresponds to the magnitude of the strain $x_n$ (top) and $y_n$ (bottom). Right panel: same simulation as in the left panel but at frequency $\o_b  = 2.10$. Here $\k =1$ and $\r = 1/3$.}
\label{fig:r03bc_evol1}
\end{figure}

In contrast to the site-centered solutions, the bond-centered ones exhibit only real instability at the early stage of the continuation when the breather frequency $\o_{b}$ is greater than but close to $\o$. At those frequencies, the magnitude of the Floquet multipliers corresponding to the real instability of bond-centered breathers is larger than the moduli of the multipliers describing the oscillatory instability of the site-centered ones, resulting in not only shorter lifetime of the bond-centered solutions, but also setting the dark breather state in motion.
A representative space-time evolution diagram for bond-centered dark breather solution of frequency $\o_b = 2.05$ is shown in the left panel of Fig.~\ref{fig:r03bc_evol}. In the right panel of Fig.~\ref{fig:r03bc_evol} the manifestation of real instability of the same solution is shown, where the perturbation of the dark breather solution along the direction associated with the unstable mode that corresponds to a real Floquet multiplier is used as the initial condition for the integration. One can see that the instability results in a dark breather moving with constant velocity after some initial transient time in the left
panel, while the pertinent motion is initiated essentially immediately
by the perturbation induced in the right panel.
However, as $\o_{b}$ is increased the same phenomenology (dismantling of the breather and chaotic evolution) is also taking place
for the bond-centered breathers, as shown in Fig.~\ref{fig:r03bc_evol1}.

The large arc seen in the middle of the Floquet multiplier diagram of the bond-centered breather solutions (the top left plot in Fig.~\ref{fig:r03floq}) corresponds to the \emph{period-doubling bifurcation}. As the frequency approaches $\o_b \approx 2.063$, two complex conjugate eigenvalues collide on the real axis at $-1$. Two real eigenvalues then form and move in  opposite directions as $\o_b$ increases. After the difference between the real eigenvalues reaches a maximum value, they start moving toward each other and collide at $\o_b \approx 2.096$. Breathers with double the period (half the frequency) of the ones on the main branch exist between these two frequencies.

\begin{figure}[htp]
\centering
\centerline{\psfig{figure=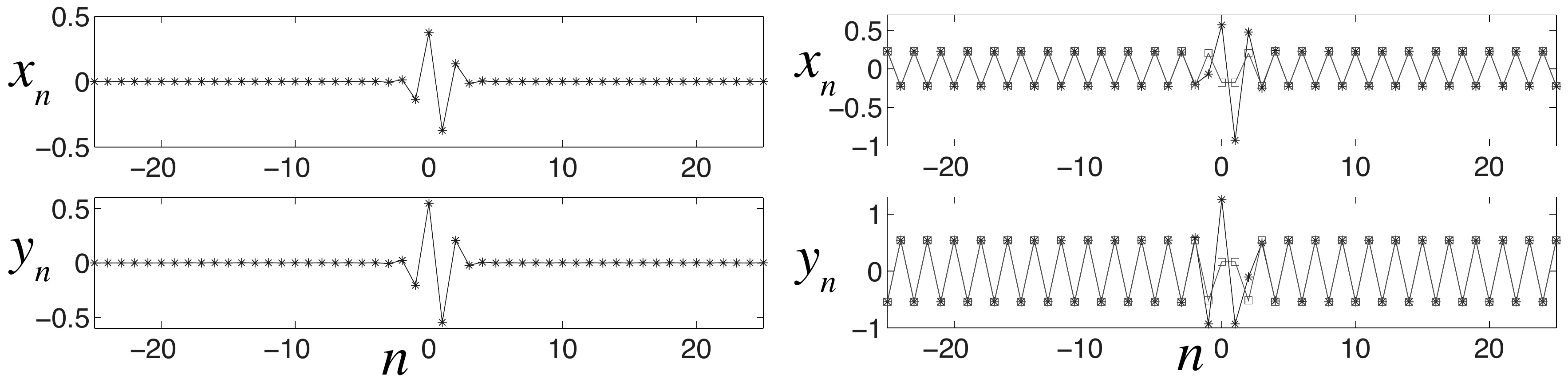,width=\textwidth}}
\caption{\footnotesize Left panel: the eigenvector associated with the eigenvalue $-1$. Right panel: numerically exact dark breather solution of frequency $\o_b = 1.032$ (connected squares) and the initial seed (connected stars) obtained by perturbing numerically exact dark breather solution of frequency $\o_b = 2.063$.}
\label{fig:PdoubleSln}
\end{figure}

\begin{figure}[htp]
\centering
\centerline{\psfig{figure=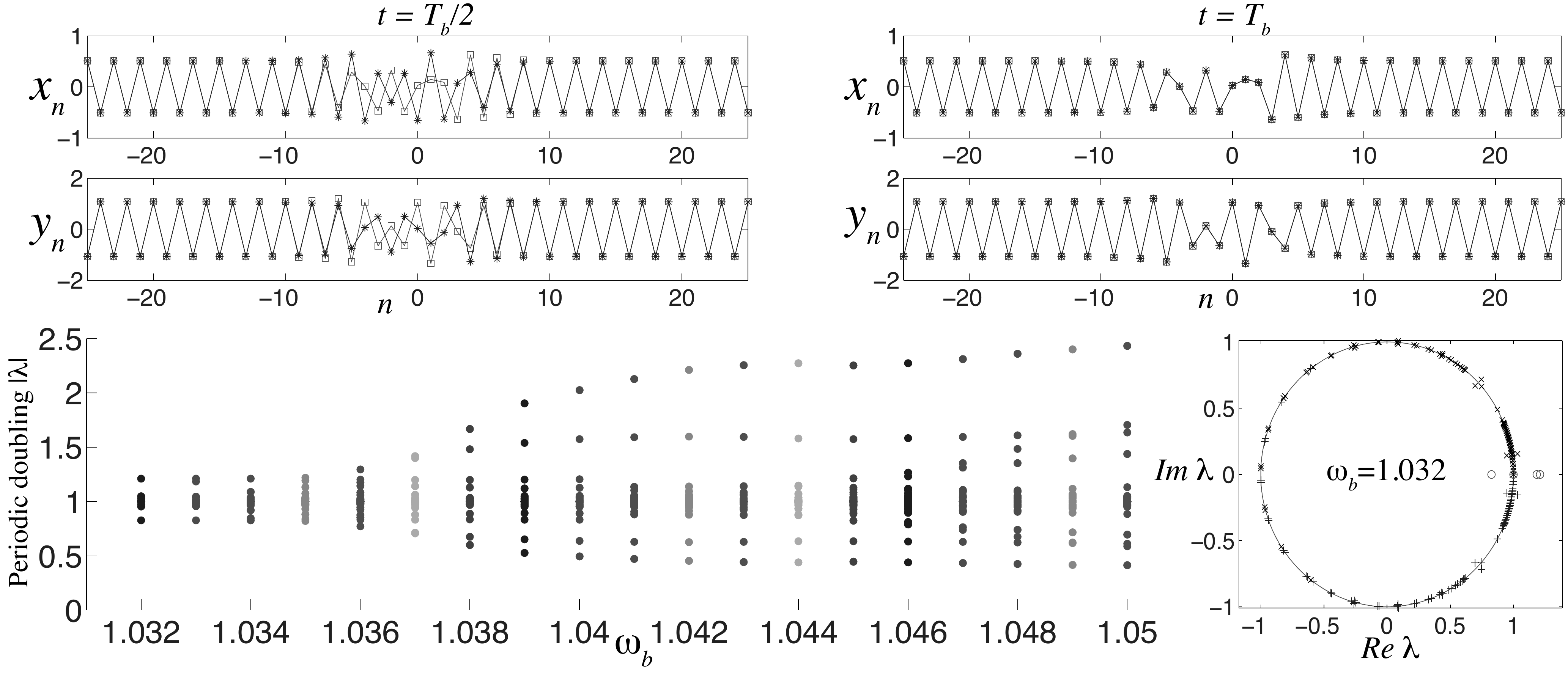,width=\textwidth}}
\caption{\footnotesize Top panel: comparison of the numerically exact double-period solution at $\o_b = 1.032$ (connected squares) and the result of its integration (connected stars) at $t={T_b}/2$ and $t=T_b$. Bottom left: moduli of Floquet multipliers versus frequency $\o_{b}$ for the period-doubling solutions. The Floquet multipliers for $\o_{b} = 1.032$ in the complex plane are shown in the right plot. Here $\k =1$ and $\r = 1/3$.}
\label{fig:PdoubleFloq}
\end{figure}

To explore the numerically exact period-doubling orbits, we constructed an initial seed by slightly perturbing the dark breather solution at the bifurcation point along the direction of eigenvector associated with the eigenvalue $-1$. As shown in Fig.~\ref{fig:PdoubleSln}, the eigenvector is spatially localized at the middle of the chain. As a consequence, the initial seed only differs from the previous dark breather solution in the middle part of the chain.
Sample profiles of period-doubling dark breather solution at $\o_b = 1.032$ are shown in the top of Fig.~\ref{fig:PdoubleFloq}. To check that these solutions differ from the main branch, we integrated the solution for both the full period $T_b = 2\pi/\o_b$ and its half $T_b/2$ and verified that the period of the obtained new solution is doubled compared with the previously obtained dark breathers. The continuation method is used to obtain all the period-doubling dark breather solutions for different frequencies. Note that the continuation stops at $\o_b \approx 1.051$ and also cannot proceed for $\o_{b}$ below $1.032$, which agrees with the (doubled) frequency range of the large arc in the top left plot of Fig.~\ref{fig:r03floq}. All of these solutions exhibit both real and oscillatory instabilities (see the bottom plots of Fig.~\ref{fig:PdoubleFloq}).

\begin{figure}[htp]
\centering
\centerline{\psfig{figure=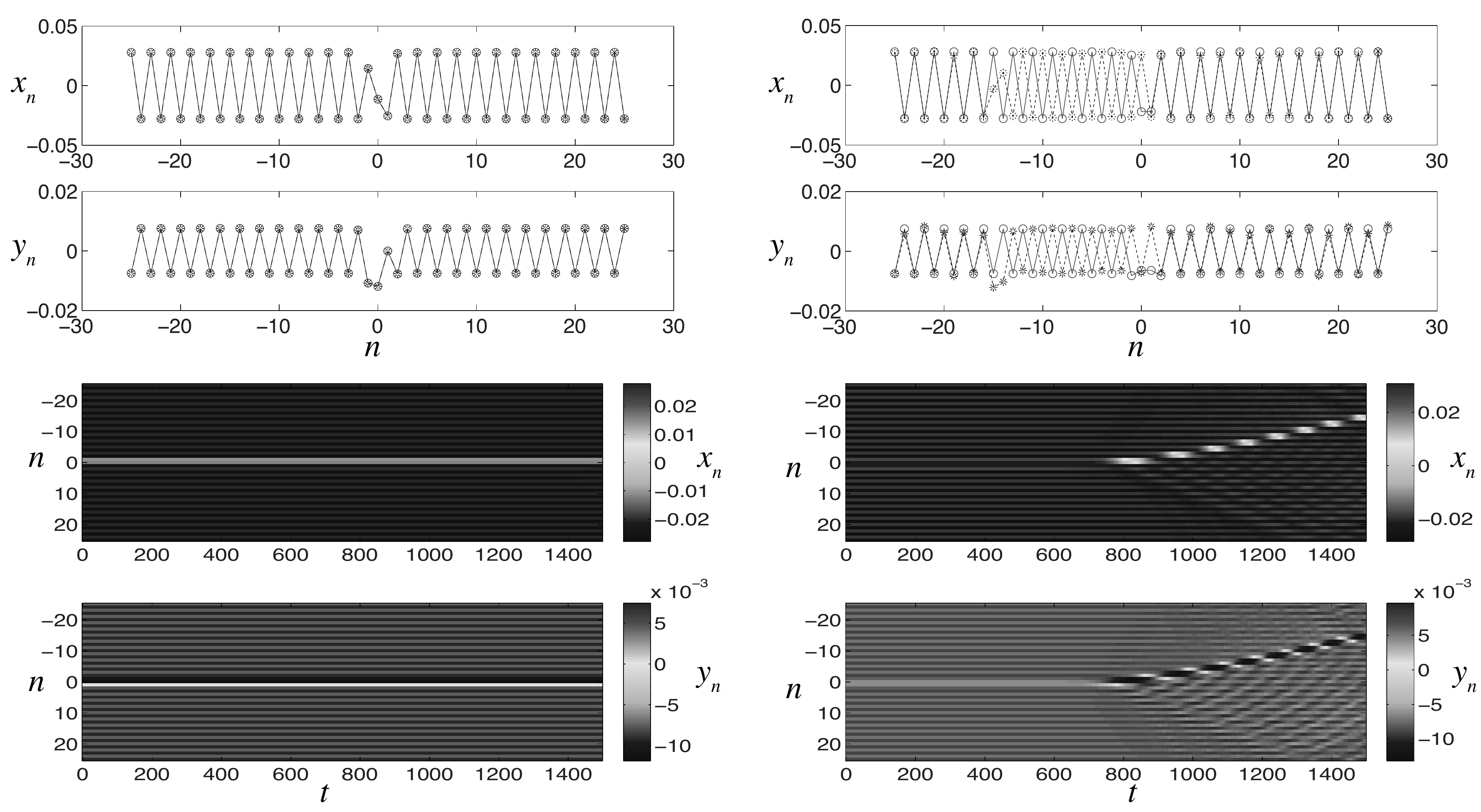,width=\textwidth}}
\caption{\footnotesize Top panel: sample profiles (circles) of site-centered (left) and bond-centered (right) dark breathers at the frequency $\o_b = 1.2547$. Stars connected by dashed lines represent strain profiles after the integration over $299T_b \approx 1500$. Note that the site-centered solution has relative error $E_b(1500) = 9.21\times10^{-5}$. Bottom panel: space-time evolution diagrams for site-centered (left) and bond-centered (right) solutions. Here $\k =1$ and $\r = 3$.}
\label{fig:r3evol}
\end{figure}

\begin{figure}[htp]
\centering
\centerline{\psfig{figure=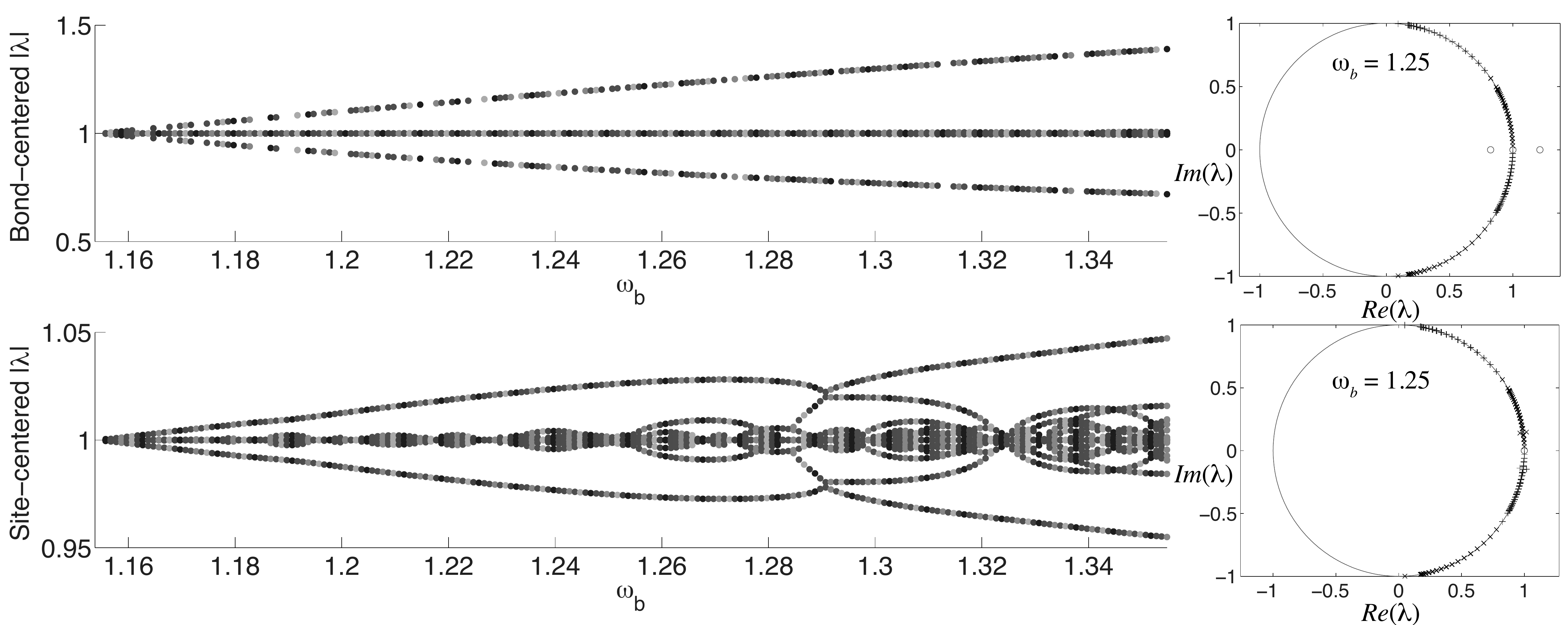,width=\textwidth}}
\caption{\footnotesize Left panel: moduli of Floquet multipliers versus frequency $\o_{b}$ for the bond-centered (top) and site-centered (bottom) types. Right panel: Floquet multipliers of dark breather solutions of frequency $\o_b = 1.2547$ in the complex plane. Here $\k =1$ and $\r = 3$.}
\label{fig:r3floq}
\end{figure}

We now investigate the effect of mass ratio $\r$ on the system by repeating the same experiment for relatively large and small values $\r$. Recall that the linear frequency $\o = \sqrt{\k + \k/\rho}$ decreases and approaches $\sqrt{\k}$ as $\r$ increases. In what follows, we start the continuation at frequency $\o_{b} = \o + 0.01$ to make sure that the amplitude of the initial seed \eqref{eq:xy_sw} is small. We first test the case when $\r = 3$ with linear frequency given by $\o = 1.1547$. Sample profiles of bond-centered and site-centered dark breather solutions at the frequency $\o_{b} = \o + 0.1 = 1.2547$ and their space-time evolution diagrams are shown in Fig.~\ref{fig:r3evol}. We observe the bond-centered solution starts to move in form of a traveling dark breather after the integration for a sufficiently long time, while the site-centered solution persists for a longer time in the simulation and hence can be considered to be long-lived. The dynamic behavior of both solutions is consistent with their numerically computed Floquet spectrum shown in the right of Fig.~\ref{fig:r3floq}. Moreover, the diagrams of Floquet multipliers' moduli in the right panel of Fig.~\ref{fig:r3floq} suggest that the bond-centered solutions exhibit only real instability for a wide range of frequencies $[\o + 0.001, \o + 0.2]$, while the site-centered dark breathers have just marginal oscillatory instability.

\begin{figure}[htp]
\centering
\centerline{\psfig{figure=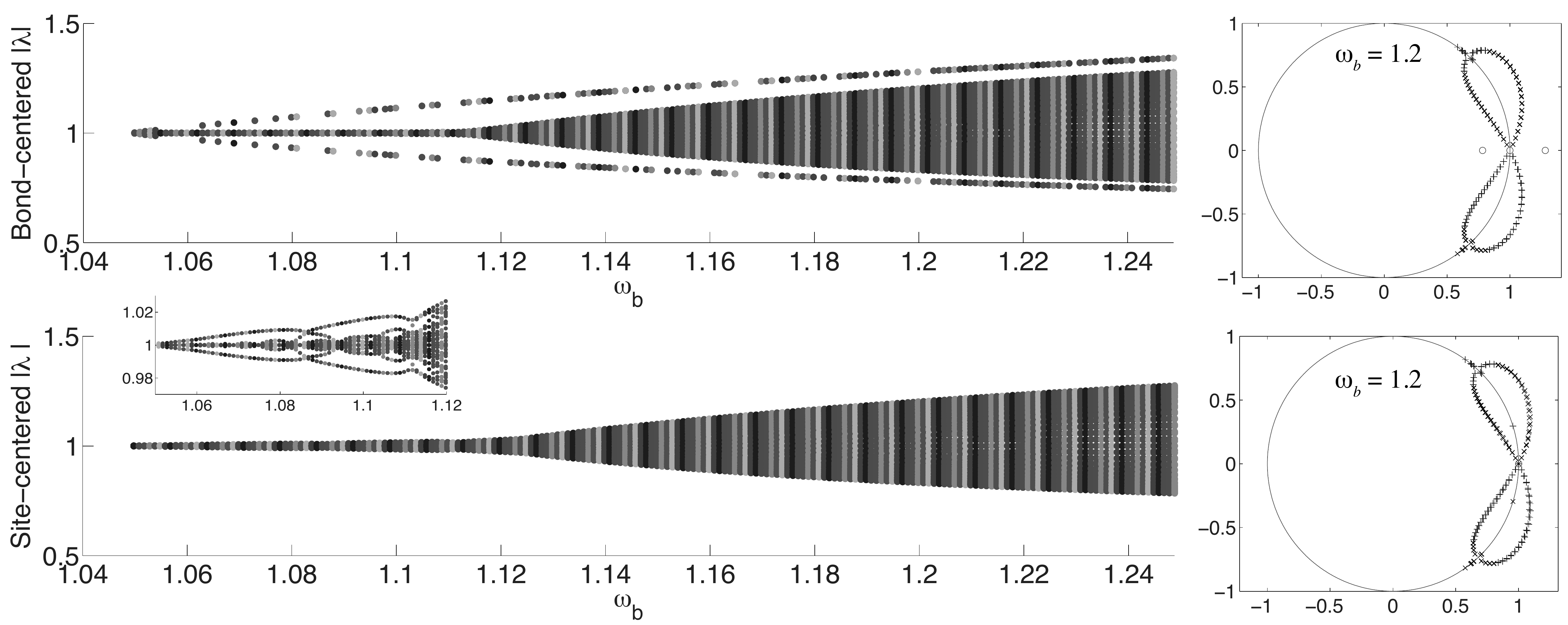,width=\textwidth}}
\caption{\footnotesize Left panel: moduli of Floquet multipliers versus frequency $\o_{b}$ for the bond-centered (top) and site-centered (bottom) types. Right panel: Floquet spectrum of dark breather solutions of frequency $\o_b = 1.20$ in the complex plane. The inset represents the zoom-in of moduli of Floquet multipliers for the site-centered solution at the frequencies $\o_b \in [w + 0.001, w + 0.07]$. Here $\k =1$ and $\r = 10$.}
\label{fig:r10floq}
\end{figure}

Next, we repeat the simulation with $\r = 10$ and $\o = 1.0488$. Note that the amplitude of dark breather solution at frequency $\o_b = \o + 0.01$ is close to $1.6\times 10^{-4}$. As shown in Fig.~\ref{fig:r10floq}, the pattern of Floquet multipliers moduli is very similar to the $\r =3$ case for breather frequencies close to $\o$. However, as $\o_b$ becomes larger, we observed significant oscillatory instability of both the bond-centered and site-centered solutions. Note that the distribution of Floquet multipliers at large mass ratios (for example, $\r = 3$, $10$) is completely different than that at smaller ones (such as $\rho = 1/3$), given the same magnitude of frequency difference $\o_b - \o$.
%{\bf The existence of these arcs of unstable multipliers suggest
% a strong (modulational) instability of the background in this setting.}

\begin{figure}[htp]
\centering
\centerline{\psfig{figure=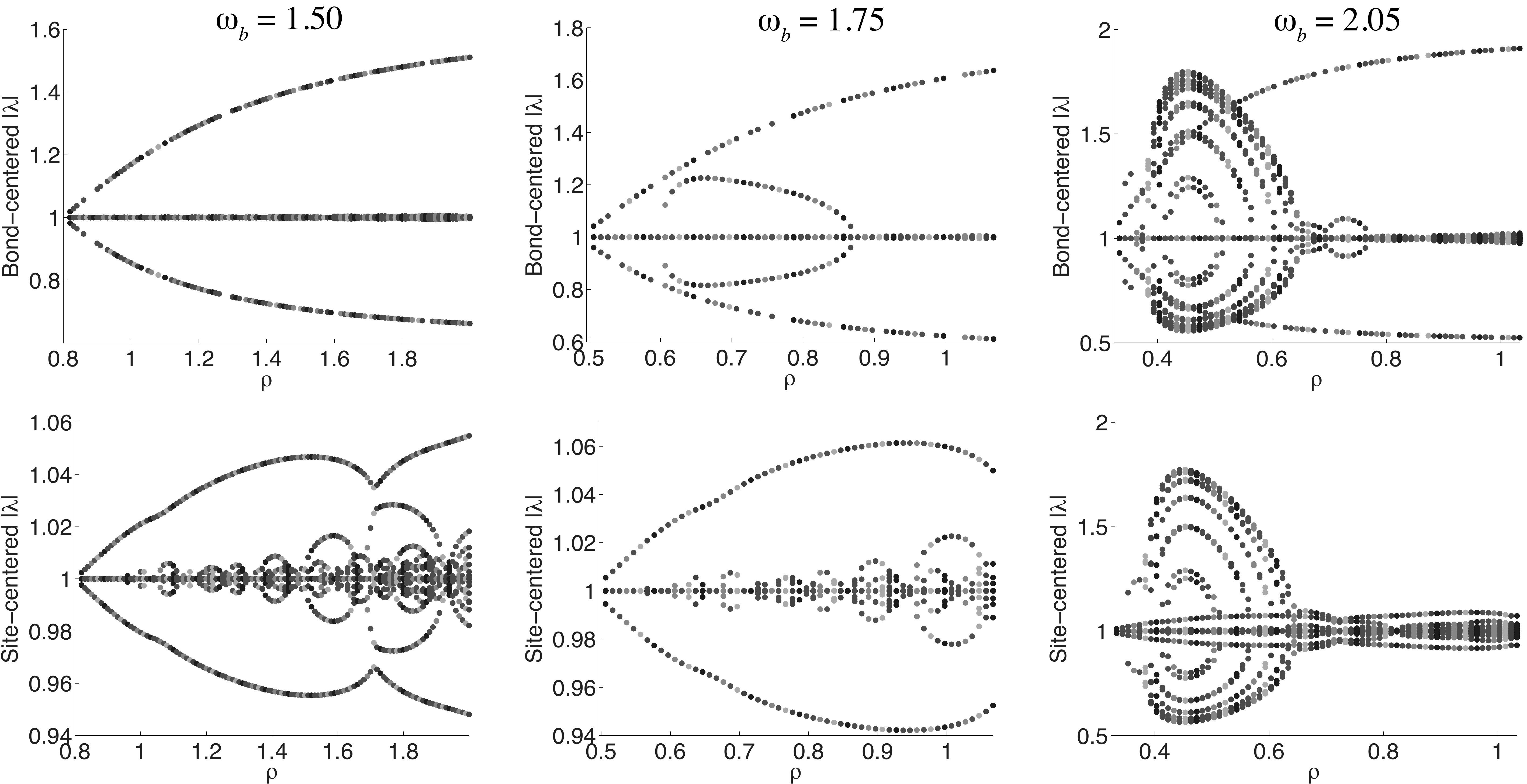,width=\textwidth}}
\caption{\footnotesize Moduli of Floquet multipliers versus mass ratio $\r$ for the bond-centered (top) and site-centered (bottom) type. The tested frequencies of the dark breather are $\o_b = 1.50, 1.75$ and $2.05$. Here $\k = 1$.}
\label{fig:rhocts}
\end{figure}

We now fix the breather frequency $\o_b$ and perform the continuation in the mass ratio $\r$. The results of numerical continuation are shown in Fig.~\ref{fig:rhocts}. At a given breather frequency, the real instability is only exhibited by solutions of the bond-centered type, and its significance is gradually increasing as mass ratio becomes larger. In contrast, the site-centered solutions have marginal oscillatory instability and persist for a long time. Moreover, the lifetime of those solutions decreases as the mass ratio increases. At a larger frequency like $\o_b = 2.05$ and small mass ratio, the emergence of many unstable quartets suggests that both bond-centered and site-centered solutions share strong modulational instabilty of the background, which leads to a chaotic evolution of both solutions after a short time of integration.
%

%%%%%%%%
\section{Concluding remarks}\label{sec:end}
In this work, we studied nonlinear waves in a resonant granular material modeled by a Hertzian chain of identical particles with a secondary mass attached to each bead in the chain by a linear spring. Following the approach developed in \cite{GJ11} for a limiting case of the present model, we derived generalized modulation equations of DpS type. We showed that for suitable initial data and large enough mass ratio, these equations reduce to the DpS equation derived in \cite{GJ11} and rigorously justified the equation in this limit on the long time scales. We then used the DpS equations to investigate the time-periodic traveling wave of the system at finite mass ratio. We showed numerically that these equations can successfully capture the dynamics of small-amplitude periodic traveling waves.

Turning our attention to the breather-type solutions, we proved non-existence of nontrivial bright breathers at finite mass ratio. However, we also showed that at sufficiently large mass ratio and suitable initial data, the problem has long-lived bright breather solutions.

The generalized DpS equations were also used to construct well-prepared initial conditions for the numerical computation of dark breather solutions.
A continuation procedure based on a Newton-type fixed point method and initiated by the approximate dark breather solutions obtained from the DpS equations was utilized to compute numerically exact dark breathers for a wide range of frequencies and at different mass ratios. The stability and the bifurcation structure of the numerically exact dark breathers of both bond-centered and site-centered types were examined. Our numerical results strongly suggest that the bond-centered solutions exhibit real instability that may give rise to steady propagation of a dark breather after large enough time. In addition, period-doubling bifurcations of these solutions were identified at small mass ratios. The site-centered solutions, in contrast to the bond-centered ones, appeared to exhibit only oscillatory instability, which is much weaker than the real instability of the bond-centered breathers for a range of breather frequencies that are close enough to the natural
frequency of the system, i.e., the frequency of out-of-phase motion
within each unit cell of the chain involving the particle and the
secondary mass. As a consequence, these low-frequency site-centered solutions persisted for a long time in the numerical simulation, and thus the effect of oscillatory instability is quite weak.
However, we also provided case examples of their (long-time)
instabilities that led to
their complete destruction and ensuing apparently chaotic dynamics within
the lattice.
We showed that the distribution of Floquet multipliers and hence stability of the dark breather solutions are significantly affected by the mass ratio and breather frequency.

A challenge left for the future work is to rigorously prove the existence of small-amplitude exact periodic traveling wave and dark breather solutions of system \eqref{eq:Hertz} using the approximate solutions obtained from the generalized DpS equations. Another intriguing aspect to further consider involves the
mobility of the dark breathers, and its association with the dynamical
instability of the states, as well as possibly with the famous
Peierls-Nabarro barrier associated with the energy difference between
bond- and site-centered solutions, i.e., the energy barrier that needs
to be ``overcome'' in order to have mobility of the dark breathers.
Equally important and relevant would be an effort to analytically understand
the modulational stability properties of the lattice, perhaps at the DpS
level and compare them with corresponding systematic numerical computations.
On the experimental side, it will be interesting to investigate whether we can generate dark breathers by exciting the both ends of a finite chain in a
way similar to~\cite{chong14}. Additionally, exciting small amplitude
traveling waves through boundary excitations, e.g. in the woodpile
chain of~\cite{Kim14} and observing experimentally their evolution
through lased Doppler vibrometry would also be particularly relevant.\\

\noindent {\it Acknowledgements.} G.J. acknowledges financial support from the Rh\^one-Alpes Complex Systems Institute (IXXI). The work of L.L. and A.V. was partially supported by the US NSF grant DMS-1007908. P.G.K. gratefully acknowledges the support
of US AFOSR
through grant FA9550-12-1-0332. P.G.K.'s work at Los Alamos is supported
in part by the US Department of Energy.

%%%%%%%%%%%%%%%%%%%%%%%%%%%%%%%%
%%%%%%%%%%%%%%%%%%%%%%%%%%%%%%%%
\bibliography{refs_notes}
\end{document}